\documentclass{pasj01}
\draft
\usepackage{color}
\usepackage{hangcaption}  
\usepackage{lscape}
\usepackage{longtable}

\Received{2016/02/12}
\Accepted{2016/07/29}
\Published{}
\begin{document}
\title{{Large} X-ray Flares on Stars Detected with MAXI/GSC: A Universal
Correlation between the Duration of a Flare and its X-ray Luminosity}

\author{Yohko \textsc{Tsuboi}\altaffilmark{1},
 Kyohei \textsc{Yamazaki}\altaffilmark{1},
 Yasuharu \textsc{Sugawara}\altaffilmark{1},
 Atsushi \textsc{Kawagoe}\altaffilmark{1},
 Soichiro \textsc{Kaneto}\altaffilmark{1},
 Ryo \textsc{Iizuka}\altaffilmark{1,2},
 Takanori \textsc{Matsumura}\altaffilmark{1},
 Satoshi \textsc{Nakahira}\altaffilmark{3},
 Masaya \textsc{Higa}\altaffilmark{1},
 Masaru \textsc{Matsuoka}\altaffilmark{3,4},
 Mutsumi \textsc{Sugizaki}\altaffilmark{3},
 Yoshihiro \textsc{Ueda}\altaffilmark{5},
 Nobuyuki \textsc{Kawai}\altaffilmark{3,6},
 Mikio \textsc{Morii}\altaffilmark{6},
 Motoko \textsc{Serino}\altaffilmark{3},
 Tatehiro \textsc{Mihara}\altaffilmark{3},
 Hiroshi \textsc{Tomida}\altaffilmark{4},
 Shiro \textsc{Ueno}\altaffilmark{4},
 Hitoshi \textsc{Negoro}\altaffilmark{7},
 Arata \textsc{Daikyuji}\altaffilmark{8},
 Ken \textsc{Ebisawa}\altaffilmark{2},
 Satoshi \textsc{Eguchi}\altaffilmark{9},
 Kazuo \textsc{Hiroi}\altaffilmark{5},
 Masaki \textsc{Ishikawa}\altaffilmark{10},
 Naoki \textsc{Isobe}\altaffilmark{11},
 Kazuyoshi \textsc{Kawasaki}\altaffilmark{4},
 Masashi \textsc{Kimura}\altaffilmark{12},
 Hiroki \textsc{Kitayama}\altaffilmark{12},
 Mitsuhiro \textsc{Kohama}\altaffilmark{4},
 Taro \textsc{Kotani}\altaffilmark{13},
 Yujin E. \textsc{Nakagawa}\altaffilmark{3},
 Motoki \textsc{Nakajima}\altaffilmark{14},
 Hiroshi \textsc{Ozawa}\altaffilmark{7},
 Megumi \textsc{Shidatsu}\altaffilmark{5},
 Tetsuya \textsc{Sootome}\altaffilmark{3,15},
 Kousuke \textsc{Sugimori}\altaffilmark{6},
 Fumitoshi \textsc{Suwa}\altaffilmark{7},
 Hiroshi \textsc{Tsunemi}\altaffilmark{12},
 Ryuichi \textsc{Usui}\altaffilmark{6},
 Takayuki \textsc{Yamamoto}\altaffilmark{3,7},
 Kazutaka \textsc{Yamaoka}\altaffilmark{13},
 and Atsumasa \textsc{Yoshida}\altaffilmark{3,13} 
}

\altaffiltext{1}{Department of Physics, Faculty of Science and
Engineering, Chuo University, 1-13-27 Kasuga, Bunkyo-ku, Tokyo 112-8551}
\email{tsuboi@phys.chuo-u.ac.jp}
\altaffiltext{2}{Institute of Space and Astronautical Science (ISAS), Japan Aerospace
Exploration Agency (JAXA), 3-1-1 Yoshino-dai, Chuo-ku, Sagamihara, Kanagawa 252-5210}
\altaffiltext{3}{MAXI team, RIKEN, 2-1 Hirosawa, Wako, Saitama 351-0198}
\altaffiltext{4}{ISS Science Project Office, Institute of Space and
Astronautical Science (ISAS), Japan Aerospace Exploration Agency (JAXA),
2-1-1 Sengen, Tsukuba, Ibaraki 305-8505}
\altaffiltext{5}{Department of Astronomy, Kyoto University, Oiwake-cho,
Sakyo-ku, Kyoto 606-8502}
\altaffiltext{6}{Department of Physics, Tokyo Institute of Technology,
2-12-1 Ookayama, Meguro-ku, Tokyo 152-8551}
\altaffiltext{7}{Department of Physics, Nihon University, 1-8-14
Kanda-Surugadai, Chiyoda-ku, Tokyo 101-8308}
\altaffiltext{8}{Department of Applied Physics, University of Miyazaki,
1-1 Gakuen Kibanadai-nishi, Miyazaki, Miyazaki 889-2192}
\altaffiltext{9}{National Astronomical Observatory of Japan, 2-21-1,
Osawa, Mitaka City, Tokyo 181-8588}
\altaffiltext{10}{School of Physical Science, Space and Astronautical
Science, The graduate University for Advanced Studies (Sokendai), 3-1-1
Yoshino-dai, Chuo-ku, Sagamihara, Kanagawa 252-5210}
\altaffiltext{11}{Institute of Space and Astronautical Science (ISAS),
Japan Aerospace Exploration Agency (JAXA), 3-1-1 Yoshino-dai, Chuo-ku,
Sagamihara, Kanagawa 252-5210}
\altaffiltext{12}{Department of Earth and Space Science, Osaka
University, 1-1 Machikaneyama, Toyonaka, Osaka 560-0043}
\altaffiltext{13}{Department of Physics and Mathematics, Aoyama Gakuin
University,5-10-1 Fuchinobe, Chuo-ku, Sagamihara, Kanagawa 252-5258}
\altaffiltext{14}{School of Dentistry at Matsudo, Nihon University,
2-870-1 Sakaecho-nishi, Matsudo, Chiba 101-8308}
\altaffiltext{15}{Department of Electronic Information Systems, Shibaura
Institute of Technology, 307 Fukasaku, Minuma-ku, Saitama, Saitama
337-8570}


\maketitle

\begin{abstract}
23 giant flares from 13 active stars (eight RS CVn systems, one Algol
system, three dMe stars and one YSO) were detected during the first two
years of our all-sky X-ray monitoring with the gas propotional counters
(GSC) of the Monitor of All-sky X-ray Image (MAXI). The observed
parameters of all of these MAXI/GSC flares are found to be at the upper
ends for stellar flares with the luminosity of $10^{31-34}$ ergs
s$^{-1}$ in the 2--20 keV band, the emission measure of 10$^{54-57}$
cm$^{-3}$, the $e$-folding time of 1 hour to 1.5 days, and the total
radiative energy released during the flare of 10$^{34-39}$ ergs.
Notably, the peak X-ray luminosity of 5$^{+4}_{-2} \times10^{33}$ ergs
s$^{-1}$ in the 2--20 keV band was detected in one of the flares on II
Peg, which is one of the, or potentially the, largest ever observed in
stellar flares.  X-ray flares were detected from GT Mus, V841 Cen, SZ
Psc, and TWA-7 for the first time in this survey.  Whereas most of our
detected sources are multiple-star systems, two of them are single stars
(YZ CMi and TWA-7).  Among the stellar sources within 100 pc distance,
the MAXI/GSC sources have larger rotation velocities than the other
sources. This suggests that the rapid rotation velocity may play a key
role in generating large flares.  Combining the X-ray flare data of
nearby stars and the sun, taken from literature and ou\ r own data, we
discovered a universal correlation of $\tau \propto L_{\rm X}^{0.2}$ for
the flare duration $\tau$ and the intrinsic X-ray luminosity $L_{\rm X}$
in the 0.1--100 keV band, which holds for 5 and 12 orders of magnitude
in $\tau$ and $L_{\rm X}$, respectively.  The MAXI/GSC sample is located
at the highest ends on the correlation.
\end{abstract}

\section{Introduction}

Cool stars, which have spectral types of F, G, K, and M, are known to
show X-ray flares. The flares are characterized with the fast-rise and
slow-decay light curve. The flares generally accompany
the rise and decay in the plasma temperature. The
general understanding, based on the numerous
studies of solar flares, is that such features  arise as a
consequence of a sudden energy release and relaxation process in the
reconnection of magnetic fields on/around stellar surfaces. In solar
flares, the reconnection, which occurred in somewhere at large coronal
heights, accelerates primarily electrons (and possibly ions) up to MeV
energies, and the electrons precipitate along the magnetic fields into
the chromosphere, suddenly heating the plasma at the bottom of the
magnetic loop to very high temperatures. A large amount of plasma
streams from the bottom to the top of the magnetic loop, while
cooling has already started by that time. The flare temperature thus peaks before the
Emission Measure ({\it EM}) does, or analogously, harder emission peaks
before softer emission.

Numerous studies on flare stars have been made with pointing
observations. For the reviews, see \citet{Pettersen89},
\citet{Haisch+91}, \citet{Favata+03}, \citet{Gudel04}, and references
therein.  However, we cannot yet answer  some fundamental
questions, such as how large a flare  a star can have, and how  very large flares
are generated. The poor understanding is rooted in the fact that the
larger flares occur less frequently.  Hence, all-sky monitoring is
crucial to detect such large flares.

X-ray all-sky monitors like Ariel-V/SSI, GRANAT/WATCH, and Swift/BAT
have  detected some large stellar flares. Using the data of
Ariel-V/SSI spanning for 5.5 years, \citet{Pye+83} and \citet{Rao+87}
detected in total twenty flares from seventeen stellar sources, including
ten RS CVn systems and seven dMe stars. \citet{Rao+87} showed that
there is a positive correlation between the bolometric luminosity and
the X-ray peak luminosity.
GRANAT/WATCH detected two X-ray transients, which have a counterpart
of  a flare star in their respective positional error boxes \citep{Castro+99}.
Swift with BAT  prompted the follow-up observations with XRT after
detecting large flares from an RS CVn system II Peg
\citep{Osten+07} and that from a dMe star EV Lac \citep{Osten+10}. 
Flares from two other RS CVn stars (CF Tuc and
UX Ari) have been detected with Swift/BAT \citep{Krimm+13}.

 Following  successful detections of large flares with all-sky X-ray
surveys, we executed a survey of stellar flares with the Monitor of
All-sky X-ray Image (MAXI; \cite{Matsuoka+09}). MAXI is a mission of
an all-sky X-ray monitor operated in the Japanese Experiment Module
(JEM; Kibo) on the International Space Station (ISS) since 2009
August. It observes an area in the sky once per 92 min orbital cycle,
and enables us to search for stellar flares effectively. In this
paper, we report the results with the gas proportional counters (GSC)
of MAXI obtained by the first two-years operation from 2009 August to
2011 August. The results with the CCD camera of MAXI (SSC) will be
given elsewhere. We describe the MAXI observation in $\S$2, our
flare-search method and the results in $\S$3, then discuss the
properties of the detected flares and the flare sources in $\S$4.

\section{Observations}

The MAXI has two types of slit cameras, the GSC and SSC, both of which
incorporate X-ray detectors consisting of gas proportional counters
and X-ray CCDs, respectively. These detectors cover an energy range of
2 to 30 keV and 0.5 to 12 keV, respectively \citep{Matsuoka+09,
  Tsunemi+10, Tomida+11, Mihara+11}. As stated in section $\S$1, the
observations of stellar flares examined here were conducted by the
GSC, which has a larger field of view (FoV) and then sky coverage than
SSC. The GSC achieves better sensitivity in the 2--10 keV band than
any other X-ray all-sky monitors so far using large-area proportional
counters with a low background, and then is preferable to detect stellar flares.
The data from 2009 August 15th to 2011 August 15th are used here.

The GSC consists of twelve pieces of proportional counters, which employ
resistive carbon-wire anodes to acquire one-dimensional position
sensitivity.  Each set of two counters forms a single camera unit, of
which the GSC has six in total. The overall FoV is a slit shape of
$\timeform{160D} \times \timeform{3D}$ in the horizon and zenith
directions, respectively, which allows the MAXI to scan the entire sky
twice as the ISS moves; i.e., MAXI/GSC can scan 97\% of the entire sky
with each ISS orbit. When the ISS passes high background regions such as
the South Atlantic Anomaly, the high voltage of the GSC is switched off
to protect the proportional counters from damage. Then, the actual sky
coverage  is about 85\% of the whole sky per 92-minute orbital period, 95\% per
day, and 100\% per week.

The point spread function (PSF) in the Anode-Wire Direction is
determined by the angular response of the slit-and-slat collimator and
the positional response of the position-sensitive gas counter along the
anode wire.  The collimator is designed to have an angular resolution of
\timeform{1D.0}--\timeform{1D.5} in FWHM, depending on the X-ray
incident angle in the Anode-Wire Direction and X-ray energy.  The PSF in
the scan direction is determined with the modulated time variations of
the detector area, which changes according to the triangular transmission
function of the collimator during each transit.  The GSC typically scans
a point source on the sky during a transit of 40--150 seconds with a FoV
of \timeform{1D.5}-width (FWHM) every 92-minute orbital period. The
transit time depends on the source incident angle in the Anode-Wire
Direction. The detector area for the target changes according to the
triangular transmission function of the collimator during each
transit. The peak value is 4--5 cm$^2$ per one camera.  The detailed
performance of the GSC was described by \citet{Sugizaki+11}.
All the data we used  were delivered from the MAXI database system
\citep{Negoro+16}.

\section{Analysis and Results}

\subsection{Search for Flaring Stars}
In order to search for flares from stars, we used the alert system
``nova search'' \citep{Negoro+10}. The alert system on the ground
swiftly reports X-ray transient events to astronomers worldwide,
prompting potential follow-up observations. For a further search, we
have created movies of GSC image for each sky area segmented into
circles of \timeform{10D} radii, setting the observation time of one day
for one shot.  The entire sky is covered with about 200 segments.

From the confirmed transient events, we selected events whose peaks are
located within \timeform{2D} from nearby known stellar sources. The
source lists are composed of the catalogs of \citet{Torres+06},
\citet{Lopez+06}, and \citet{Riedel+14}. The locations of the X-ray
peaks are determined automatically for the sources confirmed by ``nova
search'', or by eye from the movies for the others. For these selected
stellar-flare candidates, we proceeded to the following identification
process.

\subsection{Significance  in Source Detection}
We estimated the significance of a detected source  with the same
method employed in \citet{Uzawa+11}, as follows:
(1)  extracting  events within a circle with \timeform{1D.5} radius centered at the transient event (``source region'') in the 2--10 keV band,
(2)   $\sim$10 circle regions, whose
radii are all
\timeform{1D.5}, are chosen around the source region (``background circles''),
(3)  counting the numbers of events in the 2--10 keV band in  each
background circle,
(4)  the average of the counts in a background circle is defined as the background level,
(5)  the standard deviation of the counts in a background
circle is regarded as the 1-$\sigma$ background fluctuation,
(6)  subtracting the background level from the counts in the source region, and  the residual is regarded as the source counts,
(7)  the source counts divided by the 1-$\sigma$ background fluctuation is defined as the source significance.

 In the first flare on II Peg (FN15 in table~\ref{FN}) on 2009 August
20th, the source was in a high-noise area,  located close 
to the wire edge. Thus, we set the background circles in the high-noise
area in order to  estimate the appropriate level of the background. The
time-spans, for which we extracted the data, are  indicated in the light
curves with horizontal bars in figure \ref{lcurve}.

\subsection{Reconfirmation of the Source Positions}
The number of the stellar flare candidates with the significance larger
than 5-$\sigma$ level were twenty-three in total. For these flares, we
further  performed two-dimensional image fittings to obtain the precise error
regions for the X-ray positions. The fitting algorithm is given in
\citet{Morii+10}.  The shape of the region can be approximated with an
ellipse with the typical semi-major and semi-minor axes of
\timeform{0D.7} and \timeform{0D.5}, respectively, at 90\% confidence
level. We found that all the error regions still  encompass the
position of each stellar counterpart in our list, which we had seen within 
\timeform{2D} from the X-ray peak. We also confirmed that all the stellar
counterparts are listed in the ROSAT bright source catalog
\citep{Voges+99}. No other ROSAT bright-sources are in the same error
regions for all the events but one; the error region of FN20 (see
table~\ref{FN})  encompasses AT Mic (1RXS~J204151.2$-$322604) and
1RXS~J204257.5$-$320320.  1RXS~J204257.5$-$320320 is not in our list of
nearby sources, and the detailed nature is not known. Moreover,
certainly no X-ray variation has ever been reported.
Then we regard that the transient occurred on the established flare
star, AT Mic. The dates of each flare, the error regions, the X-ray
counts rates, the significant levels, and the stellar counterparts are
summarized in table~\ref{FN}.

The detected twenty-three flares were found to come from thirteen
stars; eight RS CVn systems (VY Ari, UX Ari, HR1099, GT Mus, V841 Cen,
AR Lac, SZ Psc and II Peg), one Algol-type star (Algol), three dMe
stars (AT Mic, EQ Peg and YZ CMi), and one YSO (TWA-7). Note that the
detection of the flare from TWA-7 has been already reported in
\citet{Uzawa+11}. We list the fundamental parameters of the stellar
counterparts in table~\ref{prop}. Four out of thirteen sources showed
flares  multiple times. We adopt the source distances
listed in table~\ref{prop} when we estimate  or discuss the physical parameters in this paper. The distances are all within 100 pc,
 except that of GT Mus (172 pc).  Figure \ref{Dis_Lx} displays the
relation between the source distance and $L_{\rm X}$.  This implies
that our detection limit is roughly 10 mCrab in the 2--20 keV band.

\subsection{Timing Analysis}
Figure \ref{lcurve} shows the GSC light-curves of all the detected
flares in the 2--10 keV band. In making the light-curves, the data for
the sources were extracted from the circles with the radii ranging
from \timeform{1D.3} to \timeform{1D.7}, which are selected depending
on the signal-to-noise ratio.  The backgrounds are extracted from the
annuli with the inner and outer radii of \timeform{2D} and
\timeform{4D}, respectively, except for the following two cases, in
which  an edge of a wire is close to the source region. In the case of
the flare on HR1099 (FN5) on 2010 January 23th, we chose the
background region as an annulus with the inner and outer radii of
\timeform{2D} and \timeform{3D.5}, respectively, to eliminate
a high-noise area. As for the first flare on II Peg (FN15) on 2009
August 20th, since the source was more closely located to  a wire
edge than the HR1099 case, the source region was just in the
high-noise area. Thus, in order to remove the appropriate level of the
background, we chose the background region as a rectangle of
$\timeform{3D.4} \times \timeform{19D}$, removing a central circle with
\timeform{1D.7} radius. After subtracting the background, all the
extracted source-counts were normalized by dividing by the total
exposure (in units of cm$^2$ s), which is obtained with a time integral
of the collimator effective area. We fitted them with a burst model,
which is described as a linear rise followed by an exponential
decay. The $e$-folding times are shown in table~\ref{para} and figure
\ref{Histogram}, which range from about 1 hour (AR Lac) to 1.5 days
(GT Mus).

\subsection{Spectral Analysis}
We also analyzed X-ray spectra in flare phases. The GSC spectra were
extracted during the time interval indicated with the horizontal bars
on the light-curves (figure \ref{lcurve}), as those used in the
estimation of the significance of source detection. We used the same
source and the background regions as those used in the timing
analysis. Since the photon-statistics are limited, we fitted the
spectra with a simple model; a thin-thermal plasma model ({\it mekal}:
\cite{Mewe+85,Mewe+86,Kaastra92,Liedahl+95}) with the fixed  abundance
ratios of heavy elements to the solar values. We ignored the
interstellar absorption, since all  the sources are located within
200 pc and are not in famous molecular clouds.  An example of a
spectrum with the best-fit model is found in the figure 2 in
\citet{Uzawa+11}. Table~\ref{para} and figure \ref{Histogram} give the
best-fit parameters and the distribution of the derived properties
({\it EM}, ${\it L_{\rm X}}$, $e$-folding time, the total energy),
respectively\footnote{One might  guess that significantly different results  may be
obtained with different plasma models, such as {\it apec}  \citep{Smith+01}. Then, we fitted  the spectra of FN5, FN9 and FN16 with the {\it apec} model and found that the obtained
parameters ({\it kT}, {\it EM}, ${\it L_{\rm X}}$) were consistent with
each other between the {\it mekal} and {\it apec} models for a wide range of temperature.}.

\section{Discussion}

\subsection{Detected Flares and the Source Categories}
We detected twenty-three flares, whose  X-ray luminosities are
$10^{31-34}$ ergs s$^{-1}$ in the 2--20 keV band and the emission
measures are 10$^{54-57}$ cm$^{-3}$.  The flares released the energy of 10$^{34-39}$ ergs
radiatively with the $e$-folding times of 1 hour to 1.5 days (see figure~\ref{Histogram}). All the detected flares are 
from active stars; eight RS CVn systems, one Algol system, three dMe
stars and one YSO, totaling thirteen stars.  This confirms that RS CVn
systems and dMe stars are intense flare sources, as reported in
\citet{Pye+83} and \citet{Rao+87}.  The X-ray flares from GT Mus, V841
Cen, SZ Psc, and TWA-7 were detected for the first time in this
survey. Notably, II Peg showed the $L_{\rm X}$ of 5$^{+4}_{-2}
\times~10^{33}$ ergs s$^{-1}$ in the 2--20 keV band at the peak of the
flare, which is one of the largest ever observed in the stellar flares.

Most of the flare sources that we detected with MAXI/GSC are
multiple-star system (see table~\ref{prop}).  However, two of them were
single stars: TWA-7 \citep{Uzawa+11} and YZ CMi.  In addition, another
two (AT Mic and EQ Peg) have, though a binary system, a very wide
binary-separation of roughly 6000 \RO, and so are the same as single
stars practically.  All of these four stars are known to have no
accretion disk.  These results reinforce the scenario that neither
binarity (e.g. \cite{Getman+11}) nor accretion (e.g. \cite{Kastner+02},
\cite{Argiroffi+11}), nor star-disk interaction (e.g. \cite{Hayashi+96},
\cite{Shu+97}, \cite{Montmerle+00}) is essential to generate large
flares, as has been already discussed in \citet{Uzawa+11}.

According to the catalog of active binary stars \citep{Eker+08}, 256
active binaries (e.g. RS CVn binaries, dMe binaries etc.) are known
within the distance of 100 pc from the solar system. However, we
detected flares from only ten of them. Four of them (UX Ari, HR1099, AR
Lac and II Peg) exhibited flares more than twice.

\subsection{X-ray Activity on Solar-type Stars}
As for the solar-type stars, fifteen G-type main-sequence stars are
known within the 10-pc distance (from {\it AFGK ``bright'' stars within
10
parsecs}\footnote{http://www.solstation.com/stars/pc10afgk.htm\#yellow-orange}).
The MAXI/GSC has not detected any X-ray flares from these stars. The
nearest G-type star is $\alpha$ Cen A (G2 V) at the distance of 1.3 pc
\citep{Soderhjelm99}. The upper limit on the $L_{\rm X}$ of $\alpha$ Cen
A is estimated to be 2$~\times~10^{28}$ ergs s$^{-1}$, based on the
detection limit of 10 mCrab with MAXI/GSC.  This is consistent with the
X-ray luminosities observed in solar flares; $L_{\rm X}$ is mostly lower
than $10^{27}$--$10^{28}$ ergs s$^{-1}$ \citep{Feldman+95}. However,
 large X-ray flares with the
respective $L_{\rm X}$ of $10^{29}$ ergs s$^{-1}$ and $2~\times~10^{31}$
ergs s$^{-1}$ have been observed from ordinary solar-type stars $\pi^1$ UMa \citep{Landini+86} and BD$+$10$^{\circ}$2783 \citep{Schaefer+00}. \citet{Schaefer+00} called such flares
``superflares''.  So far, very few extensive studies have been made in
the X-ray band and reported to give any good constraints on the frequency of
the occurrence of ``superflares'' in the band.  Our MAXI/GSC two-year
survey is the best X-ray study of this kind.  From our result, we can
claim that the flares with $L_{\rm X}$ of larger than $1~\times~10^{30}$
ergs s$^{-1}$ must be very rare for solar-type stars.

\subsection{$EM$ vs. $kT$ and the Derived Loop Parameters}
Figure \ref{kT_EM} shows a plot of $EM$ vs. plasma temperature $(kT)$
for the flares in our study of the MAXI/GSC sources, together with solar flares
\citep{Feldman+95}, solar microflares \citep{Shimizu95}, and flares from
the stars in literature (see table~\ref{ref_fig} for the complete set of 
references).  All of the plotted samples are roughly on the universal
correlation over orders of magnitude (\cite{Feldman+95},
\cite{Shibata+99}).  Our sample is located at the high ends in the
correlation for both the temperatures and emission measures.

Now, we consider the two important physical parameters of flares, that
is, the size and magnetic field.  \citet{Shibata+99}\ formulated the
theoretical $EM$-$kT$ relations for a given set of a loop-length and
magnetic field as equations (5) and (6) in their paper\footnote{ Their
calculation of the $EM$-$kT$ relations is based on the
magnetohydrodynamic numerical simulations of the reconnection by
\citet{Yokoyama+98}. The simulation takes account of heat conduction and
chromospheric evaporation on the following four assumptions: (1) the
plasma volume is equal to the cube of the loop length, (2) the gas
pressure of the confined plasma in the loop is equal to the magnetic
pressure of the reconnected loop, (3) the observed temperature at the
flare peak is one-third of the maximum temperature at the flare onset,
(4) the pre-flare proton ($=$ electron) number density outside the flare
loop is $10^{9}$ cm$^{-3}$.} (see figure~\ref{kT_EM} for a few
representative cases).  We calculated the loop-length and magnetic-field
strength for each of the observed flares with MAXI-GSC, based on the
relations \citep{Shibata+99}, as listed in table~\ref{para}.  The
magnetic field of our sample is comparable with those of flares on the
Sun ($\sim$15--150~G).  On the other hand, our sample has orders of
magnitude larger sizes of flare loops   than those on the Sun
($<$0.1~\RO). Especially noteworthy ones among our sample are the two largest flares
relative to their binary separations, FN4 from UX Ari and FN16 from II
Peg.  Their loop lengths are 10  and 20 times
larger than their respective binary separations, which are
unprecedentedly large among stellar flares.

The extraordinary large loop lengths could possibly be an artifact due
to the systematic error in the model by \citet{Shibata+99}.  In fact,
their derived loop lengths are 10 times larger than those obtained by
\citet{Favata+01}, who used a hydrodynamic model by \citet{Reale+98}.
In \citet{Shibata+02}, which is the follow-up paper of
\citet{Shibata+99}, it is argued that their derived loop length could be
reduced to roughly 1/10 if the two assumptions (the points 3 and 4
mentioned in the footnote below) are altered.  In our MAXI sample, even
if the true loop sizes are 1/10 of the above-estimated values as a conservative case, the
largest ones are 0.2--5 times larger than their binary separation and so
are still large.

\subsection{Duration vs. X-ray Luminosity}
We search for potential correlations in various plots to study what are
the deciding factors to generate large stellar flares and to which
extent. Figure~\ref{Decay_Lx} plots the duration of flares ($\tau_{\rm
lc}$) vs. the intrinsic X-ray luminosity ($L_{\rm X\_bol}$) in the 0.1--100 keV band for the stars detected with MAXI/GSC and with
other missions (see table~\ref{ref_fig} for the complete set of
references). Here, we have introduced $L_{\rm X\_bol}$ in order to take all the
radiative energy into our calculation \footnote{See Appendix for the detailed process
to derive $L_{\rm X\_bol}$ for each data-set.}. Solar
flares (\cite{Pallavicini+77}, \cite{Shimizu95}, \cite{Veronig+02}) too
are superposed \footnote{As the duration, we used $e$-folding time for
each flare in our work and the works introduced in
table~\ref{ref_fig}. For the data of \citet{Pallavicini+77} and that of
\citet{Veronig+02}, we used ``decay time'', of which the definitions are
up to the corresponding authors. For the data of \citet{Shimizu95}, we
used the duration itself in FWHM reported in the paper.  Generally,
flares in any magnitude have  fast-rise and slow-decay light curves, and
the rise times at longest are comparable to the corresponding decay times 
(e.g. \cite{Pallavicini+77}, \cite{Imanishi+03}).  Therefore the samples
are consistent with  one another within a factor of 2, or 0.3 in the
 logarithmic scale as in the vertical axis of figure~\ref{Decay_Lx}.}. The
data points of the MAXI/GSC flares are found to be located at the
highest ends in both the $L_{\rm X\_bol}$ and duration axes among all
the stellar flares. The plot indicates that there is a universal
correlation between $L_{\rm X\_bol}$ of a flare and its duration, such
that a longer duration means a higher $L_{\rm X\_bol}$.  Remarkably, the
correlation holds for  wide ranges of parameter values for
$10^{22}$$\lesssim L_{\rm X\_bol} \lesssim 10^{34}$~ergs~s$^{-1}$ and
$10^{1}$ $\lesssim \tau_{\rm lc} \lesssim 10^{6}$~s.  Using the datasets
of the stellar flares detected with MAXI (this work) and other missions
(table~\ref{ref_fig}), and the solar flares reported by
\citet{Pallavicini+77}, we fitted the data with a linear function in the
log-log plot \footnote{ \citet{Shimizu95} and \citet{Veronig+02}
presented the plots which indicate the X-ray luminosity and duration of
each flare, but not tables for them to give the exact values. Then we
excluded both datasets from the fitting.} and obtained the best-fit
function of
\begin{eqnarray}
\tau_{\rm lc} = (1.1^{+4.7}_{-0.9}) \times 10^{4}~\left(\frac{L_{\rm X\_bol}}{10^{33}~{\rm
				   ergs~s^{-1}}}\right)^{0.20\pm0.03}
{\rm sec},
\label{eq0}
\end{eqnarray}
 where the errors of both the coefficient and the power are in
1-$\sigma$ confidence level. The best-fit model is shown with a solid
line in figure~\ref{Decay_Lx} top panel.  We found that the best-fit
model agrees also with the range of the data for solar microflares
reported by \citet{Shimizu95}, even though the luminosities $L_{\rm
X\_bol}$ of their data are smaller than $\sim10^{25}$ ergs~s$^{-1}$, whereas
those used for our fitting are larger than that.

For comparison, 
 \citet{Veronig+02} and \citet{Christe+08} have derived similar
 power-law slope as ours, $\sim$0.33 for the GOES data and $\sim$0.2 for
 the RESSI data, respectively, though with the limited energy bands. The ranges of their luminosities  are $L_{\rm
 X} = 10^{23.5-25.5}$~ergs~s$^{-1}$ in the 3.1--24.8 keV band and
 $L_{\rm X} = 10^{22.5-25.5}$~ergs~s$^{-1}$ in the 6--12 keV band,
 respectively.  \medskip

\smallskip \indent 
In the following subsections, we discuss the
plausible models to explain this positive correlation, examining three
potentially viable scenarios: the radiative-cooling dominant,
conductive-cooling dominant, and propagating-flare models. Note that
we have chosen the former two models for simplicity and examine them
separately, although it is expected, as most star-flare
models assume, that both radiation and conduction are present in a
flare and that the latter is active early in the decay and the former
is, later (e.g. \cite{Shibata+02}, \cite{Cargill04}, \cite{Reale07}).
We assume that the duration of a flare $\tau_{\rm lc}$ represents the
cooling time of the heated plasma when we examine the radiative- and
conductive-cooling dominant models.

\subsubsection{Radiative Cooling Model}
 First, we consider the condition where the radiative cooling is
dominant. Since the
thermal energy is lost  via radiation, $\tau_{\rm lc}$ and  the
radiative cooling time $\tau_{\rm rad}$  are given by,
\begin{eqnarray}
\tau_{\rm lc} \simeq \tau_{\rm rad} 
& = &\frac{3 n_{\rm e} kT}{n_{\rm e}^{2}~F(T)}
= \frac{3 kT}{n_{\rm e}~F(T)},
\label{eq1-1}
\end{eqnarray}
where $n_{\rm e}$ and $F(T)$ are the electron density and radiative loss
rate, respectively.  We obtained the radiative loss rate, using the {\it
CHIANTI} atomic database (version 8.0) and the {\it ChiantiPy} package
(version 0.6.4) \footnote{http://www.chiantidatabase.org/chianti.html}.

On the other hand,  based on the plot of $EM$ vs. $kT$ (figure~\ref{kT_EM}),
we confirm that  most of the observed data of  stellar and solar
flares are confined in the region of 15~G $< B <$ 150~G, where $B$ is
magnetic field strength. The region is mathematically described as
\begin{eqnarray}
EM \simeq 10^{48}~
\alpha^{-5}
\left(\frac{T}{10^{7}~{\rm K}}\right)^{17/2}
~{\rm cm^{-3}}~,
\label{eq1-2}
\end{eqnarray}
where $\alpha$ is a non-dimensional parameter  ranging between 0.3 and
3. 
Since the derived $EM$ and $F(T)$ compose $L_{\rm X\_bol}$ as 
\begin{eqnarray}
L_{\rm X\_bol} = EM~F(T)~,
\label{eq1-3}
\end{eqnarray}
 we obtain,  combining it with equation (\ref{eq1-2}), 
\begin{eqnarray}
L_{\rm X\_bol} \simeq 
10^{48}~
\alpha^{-5}
\left(\frac{T}{10^{7}~{\rm K}}\right)^{17/2}~F(T)~.
\label{eq1-4}
\end{eqnarray}
Now that both $\tau_{\rm lc}$ and $L_{\rm X\_bol}$ have been parameterized
with the temperature $T$, we insert
solid lines in figure \ref{Decay_Lx} as their relation
for radiative-cooling dominant model.

Observationally, the electron density $n_{\rm e}$   was  measured to
be $10^{11}~{\rm cm^{-3}}$ in flares of Proxima Centauri with a high-resolution spectroscopy in the
X-ray band \citep{Gudel+02}.  In the solar flares, $n_{\rm e}$ of
$10^{10-11}~{\rm cm^{-3}}$ has been calculated from the $EM$ and the
volume of the loop \citep{Pallavicini+77}. 
Some other spectroscopic observations of solar flares have indicated
a wide range of $n_{\rm e}$; $10^{11-13}~{\rm cm^{-3}}$ (see references in
\cite{Gudel04}). Therefore, we derived the permitted ranges for
radiative cooling plasma  in the following two cases: (1) $n_{\rm
e}~=~10^{11}~{\rm cm^{-3}}$ and $\alpha$=~0.3--3 (figure~\ref{Decay_Lx}
middle panel), (2) $n_{\rm
e}~=~10^{10-13}~{\rm cm^{-3}}$ and $\alpha$~=~1 (figure~\ref{Decay_Lx}
bottom panel). With the wide permitted range,
the figures indicate that radiative
cooling explains all the flares in our dataset.

\subsubsection{Conductive Cooling Model}
 Second, we consider the condition where the conductive cooling is
dominant,  assuming a
semicircular loop  of the flare with the cross-section of
$\pi\left(\frac{l}{10}\right)^{2}$ for the half-length $l$,  which is often observed in solar flares.
In this case, $\tau_{\rm lc}$ and
the conductive cooling time $\tau_{\rm con}$  are given by
\begin{eqnarray}
\tau_{\rm lc} \simeq \tau_{\rm con} 
& = & \frac{3 n_{\rm e} kT}{10^{-6}~T^{7/2}~l^{-2}} \nonumber \\
& = &1.3\times10^{2}~
\left(\frac{n_{\rm e}}{10^{11}~{\rm cm^{-3}}}\right)\nonumber\\
&\times&\left(\frac{T}{10^{7}~{\rm K}}\right)^{-5/2}
 \left(\frac{l}{10^{9}~{\rm cm}}\right)^{2}
{\rm sec}.
\label{eq2-1}
\end{eqnarray}

On the other hand, a fraction of the thermal energy is observed as
radiation. The luminosity $L_{\rm X\_bol}$ and $EM$  for the temperature  $T$ are
written as  equations (\ref{eq1-3}) and (\ref{eq1-2}),
respectively. With the plasma volume of $\frac{\pi}{50}l^{3}$,
 $EM$  is also written as,
\begin{eqnarray}
EM = \frac{\pi}{50}~n_{\rm e}^{2}~l^{3}~.
\label{eq2-1-2}
\end{eqnarray}
Combining  equations (\ref{eq1-2}),
(\ref{eq2-1}) and (\ref{eq2-1-2}) gives
\begin{eqnarray}
\tau_{\rm lc} \simeq 
1.3\times10^{5}\left(\frac{50}{\pi}\right)^{2/3}~\alpha^{-10/3}~n_{\rm e}^{-1/3}~
\left(\frac{T}{10^{7}~{\rm K}}\right)^{19/6}
~.
\label{eq2-2}
\end{eqnarray}
Now that both $\tau_{\rm lc}$ and $L_{\rm X\_bol}$ have been parameterized
with the temperature $T$, we insert
dotted lines in figure \ref{Decay_Lx} as their relation
for conductive-cooling dominant model.
The values of $n_{\rm e}$ and $\alpha$ are varied 
independently  within the ranges of $10^{10}$ $\lesssim n_{\rm e} \lesssim$ $10^{13}~{\rm
cm^{-3}}$ and 0.3 $\lesssim \alpha \lesssim$ 3, respectively.
The figure indicates that  the model and the data  overlap with each other in the given parameter space.

\subsubsection{Propagating Flare Model}
 Third, we examine spatially propagating flare, like
two-ribbon flares seen on the Sun.  The total energy released during a
flare  via radiation, $E_{\rm rad}$,  is described as
\begin{eqnarray}
E_{\rm rad} \simeq \tau_{\rm lc} L_{\rm X\_bol}~.
\label{eq3-1}
\end{eqnarray}
On the other hand, $E_{\rm rad}$ originates from thermal energy 
confined in the plasma, and the thermal energy comes from stored magnetic
energy, $E_{\rm mag}$. Then,
it can be also written as 
\begin{eqnarray}
E_{\rm rad} = f~E_{\rm mag} = \frac{f~B^2~D^3}{8\pi}~,
\label{eq3-2}
\end{eqnarray}
 where $f$ is the energy conversion efficiency from magnetic energy to
that released as radiation, and $D$ is the scale length of the flaring
region.  When we assume that the magnetic reconnection propagates with
the speed $v$, $\tau_{\rm lc}$ satisfies the following formula
\begin{eqnarray}
\tau_{\rm lc} = D/v~.
\label{eq3-3}
\end{eqnarray}
Eliminating $E_{\rm rad}$ and $D$ with the equations (\ref{eq3-1}), (\ref{eq3-2}) and (\ref{eq3-3}), we obtain 
\begin{eqnarray}
\tau_{\rm lc} & \simeq & 1.2 \times 10^4 \left(\frac{f}{0.1}\right)^{1/2}
\left(\frac{B}{50~{\rm G}}\right)^{-1}\nonumber\\
& \times &\left(\frac{v}{3\times10^7~{\rm cm~s^{-1}}}\right)^{-3/2}
\left(\frac{L_{\rm X\_bol}}{10^{33}~{\rm ergs~s^{-1}}}\right)^{1/2}
 {\rm sec}.
\label{eq3-4}
\end{eqnarray}
If the values of $f$, $B$ and $v$ are common among stars, 
\begin{eqnarray}
\tau_{\rm lc} \propto  L_{\rm X\_bol}^{1/2}~.
\label{eq3-5}
\end{eqnarray}
The power of $L_{\rm X\_bol}$ is slightly larger than  what we obtained.

\subsection{Origin of  Large Flares}

\subsubsection{Rotation Velocity}
The positive correlation between quiescent X-ray luminosity and rotation
velocity has been reported by \citet{Pallavicini89} and in subsequent
studies. However, no studies have been published about this type of correlation for the flare
luminosity.  Compiling our data sample and those in
literature, we search for the potential correlation of this kind.
\newline \indent We plot the total energy released during  a flare
(i.e., $E_{\rm rad}$) vs. the square of rotation velocity ($v_{\rm
rot}^{2}$) in figure \ref{V_Etot}, where $E_{\rm rad}$ is derived by
multiplying  $L_{\rm X}$ in the 2--20 keV band by the $e$-folding
time of the flare decaying phase\footnote{ If flares have been detected
from a source multiple times with MAXI and/or other missions, only the
largest total energy is used. If the flare source is a multiple-star
system, we use the rotation velocity of the star with the largest
stellar radius in the system. In our sample, the source with the largest
radius has a higher velocity than the other stars in the same system for
all the multiple-star systems except EQ Peg.}. From figure \ref{V_Etot},
we find that the MAXI sample is concentrated at the region of the high
rotation velocity and the large total energy. This is the first
indication with an unbiased survey that stellar sources with higher
rotation velocities can have a very high $E_{\rm rad}$.

In order to validate this tendency further, we made two histograms of
the number of sources as a function of the star-rotation velocity for
flare sources.  One is for our MAXI/GSC sample, and the other is for
cataloged nearby stars from the literature (active binaries;
\cite{Eker+08}, X-ray-detected stellar sources; \cite{Wright+11}). Both
the samples are within 100~pc distance, and from the latter sample, MAXI/GSC
sources are excluded. Figure~\ref{Histogram_V} shows the two histograms. The MAXI/GSC sources have the median
logarithmic rotation velocity of 1.52 in units of log (km s$^{-1}$) with
the standard deviation of 0.37 dex.  On the other hand, the undetected
sources with MAXI/GSC have the median logarithmic value of 0.80 in the
same units with the standard deviation of 0.52 dex. Therefore, the
rotation velocities of the MAXI sources, which have shown huge flares as
reported  in this paper, are significantly higher than  those of the
other active stars that are comparatively quiet. This supports that the sources
with faster velocities generate larger flares.

\subsubsection{Stellar Radius}
We investigate whether the MAXI sources and/or the size of the
flare-emitting region have any common characteristics or correlations in their
stellar radii. \newline \indent Figure~\ref{R_EM} plots $EM$ vs. the square of
the stellar radius ($R_*^{2}$) for the stars detected with
MAXI/GSC and those detected with other missions (see table~\ref{ref_fig}
for the complete set of references)\footnote{The largest $EM$ is used
for each source if flares have been detected multiple times, and the
radius of the larger star is used if the source is a multiple-star
system.}.  Though the sample is limited, a hint of the positive
correlation between $EM$ and $R_*^{2}$ is seen at least in the MAXI/GSC
sources\footnote{ A similar plot has been made by \citet{Rao+87}, for
the flares detected with Ariel-V/SSI,  which shows a similar correlation to ours,  although  their plot was the bolometric
luminosity vs. X-ray peak luminosity.}.

We also made a number-distribution histogram as a function of the radius
in figure~\ref{Histogram_R}, similar to figure~\ref{Histogram_V}.  From
figure~\ref{Histogram_R}, we find MAXI/GSC sources are concentrated at
the radii of about 3 and 0.3 \RO, which correspond to RS CVn-type and
dMe stars, respectively.  On the other hand, the undetected sources with
MAXI/GSC are widely distributed in figure~\ref{Histogram_R}, which
implies that the magnitude of flares is not as sensitive to the stellar
radius as to the rotation velocity.

\section{Summary}

\begin{enumerate}
\item During the two-year MAXI/GSC survey, we detected twenty-three 
 energetic flares from thirteen active stars (eight RS-CVn stars, three
 dMe stars, one YSO, and one Algol type star). The physical parameters of the flares  are very large for stellar flares in all of the followings: the X-ray
 luminosity $L_{\rm X}$ ($10^{31-34}$ ergs s$^{-1}$ in the 2--20 keV
 band), the emission measure $EM$ (10$^{54-57}$ cm$^{-3}$), the
 $e$-folding time (10$^{3-6}$ s), and the total energy released during
 the flare (10$^{34-39}$ ergs).
\item
The flares from GT Mus, V841 Cen, SZ Psc
and TWA-7  were detected for the first time in the X-ray band. From II Peg, we
detected one of the largest flares among stellar flares with 
$L_{\rm X}$ of 5$^{+4}_{-2}~\times~10^{33}$ ergs~s$^{-1}$ in the
2--20 keV band.  Whereas most of the sources detected with MAXI/GSC
are multiple-star systems, two of them (YZ CMi and TWA-7) are single, which
 are known to have no accretion disk. These results
reinforce the scenario that  none of binarity,  accretion,  and
star-disk interaction is essential to generate large flares, as has
been already discussed in \citet{Uzawa+11}.
\item
The survey showed that the number of the sources that show extremely
large flares is very limited; only ten out of the 256 active binaries within the
100 pc distance  have been detected, while four of the ten sources showed 
flares  multiple times. We detected no X-ray flares from solar-type
stars, despite the fact that fifteen G-type main-sequence stars lie within 10-pc
distance. This  implies that the frequency of the superflares from solar-type stars,
which has $L_{\rm X}$ of more than $1~\times~10^{30}$ ergs s$^{-1}$, is
 very small.
\item On the $EM$--$kT$ plot, our sample is located at the high ends in
the universal correlation, which ranges over orders of magnitude
(\cite{Feldman+95}, \cite{Shibata+99}). According to the theory of
\citet{Shibata+99}, our sample has the similar intensity of magnetic
field to those detected on the Sun ($\sim$15--150 G), but has orders of
magnitude larger flare-loop sizes than those on the Sun ($<$ 0.1
\RO). The largest two loop sizes from UX Ari and II Peg are huge, and
are much larger than the binary separations.
\item 
We plotted the duration vs $L_{\rm X\_bol}$,  using the data of
solar and stellar flares in literature and the data of the flares on
MAXI/GSC sources.  The plot indicates that there is a universal positive 
correlation between  $L_{\rm X\_bol}$ of a flare and its duration, 
such that a longer duration   means a higher $L_{\rm X\_bol}$. The
correlation holds for the wide range of parameter values; 12 and 5
orders of magnitude in $L_{\rm X\_bol}$ and duration,  respectively.  Our
sample is located at the  highest ends on the correlation.  From the data,
we found that the duration  is proportional to $L_{\rm
X\_bol}^{0.2}$. 
\item
Our sample has especially fast rotation velocities with an order of 10
km s$^{-1}$. This indicates that the rotation velocity is an
essential parameter to generate big flares.
\end{enumerate}

We thank M. Sakano, Y. Maeda, F. Reale, and S. Takasao for useful discussion.
This research has made use of the MAXI
data\footnote{http://maxi.riken.jp/top/index.php}, provided by the
RIKEN, JAXA, and MAXI teams. This research was partially supported by
the Ministry of Education, Culture, Sports, Science and Technology
(MEXT), Grant-in-Aid No.19047001, 20041008, 20244015, 20540230,
20540237, 21340043, 21740140, 22740120, 23540269, 16K17667, and Global-COE from
MEXT ``The Next Generation of Physics, Spun from Universality and
Emergence'' and ``Nanoscience and Quantum Physics''. Y.T. acknowledges
financial support by a Chuo University Grant for Special Research.

\appendix
\section*{The process to derive $L_{\rm X\_bol}$ for each dataset}
In \S 4.4, we derived the intrinsic X-ray luminosity $L_{\rm X\_bol}$ in the 0.1--100 keV band
from the available observed X-ray data in narrower energy bands (\cite{Pallavicini+77}, \cite{Shimizu95}, \cite{Veronig+02}).
The process to derive  it is as follows.

\begin{enumerate}
\item
In principle, we derive $L_{\rm X\_bol}$ from the parameters $kT$ and $EM$,
using the equation~\ref{eq1-3}. As for the X-ray data of \citet{Shimizu95} and \citet{Veronig+02},
 it is necessary to estimate the values of $kT$ and $EM$
from the X-ray luminosities in the GOES band (3.1--24.8 keV) (hereafter $L_{\rm X\_GOES}$).
To estimate them, first, we derive the
ratios (denoted as $P(T)$) of the X-ray luminosity in the GOES band
to that in the 0.1--100 keV band (i.e., $L_{\rm X\_bol}$) 
as a function of $kT$, using 
the $apec$ model in $XSPEC$. 
Next, assuming the empirical relation of $EM$ vs.$kT$ for flares,  
(equation~\ref{eq1-2}), we obtain $EM$ as a function of $kT$,
 with the parameter $\alpha$ fixed to one (hereafter $EM(T)$).
 Combining these parameters with equation~4, the GOES band luminosity is written as
\begin{eqnarray}
L_{\rm X\_GOES} = L_{\rm X\_bol}~P(T) = F(T)~EM(T)~P(T).
\label{eqAppendix}
\end{eqnarray}
We  can
derive $T$ for each $L_{\rm X\_GOES}$,
and then obtain $L_{\rm X\_bol}$ with this formula.

\item As for the X-ray data of \citet{Pallavicini+90a}, we simply
  accept the X-ray luminosities in the 0.05--2 keV band in their paper
  as $L_{\rm X\_bol}$, because the luminosities in the band are
  estimated to be about 95\% of $L_{\rm X\_bol}$.  Note that we have
  used the $apec$ model in $XSPEC$, varying the temperatures in the
  range of log $T$ = 6.5--7.3 (K), to get the ratio of 95\%, where the
  range for the temperatures is assumed by \citet{Pallavicini+90a}.

\end{enumerate}

\begin{landscape}
\begin{table*}[h]
\caption{Date and  other observed parameters of each flare.}
\begin{center}
{\scriptsize
\begin{tabular}{lccccccccccl}
\hline \hline
Flare & MJD\footnotemark[$*$] & UT\footnotemark[$*$] & \multicolumn{4}{c}{Error Ellipse} &  Count Rate\footnotemark[$\dagger$] & Significance & Counterpart&Category\\
\cline{4-7}
& & &Center(J2000) &Semimajor axis&Semiminor axis&Roll angle & & & \\
 & &[YYYY MMM DD HH:MM:SS]&[HH:MM:SS, DDD:MM:SS]&[degree]&[degree]
                     &[degree]&[10$^{-2}$counts s$^{-1}$]&[$\sigma$]& &\\\hline\\
 FN1 &55144.347&2009 Nov 09 08:19:15& 02:49:55.49, +31:18:32.34
	     &1.0&0.82&90& 1.1 &7.4&VY Ari&RS CVn\\
 FN2 &55080.193&2009 Sep 06 04:37:45& 03:26:33.38, +28:31:36.03 &0.6&0.57&90& 0.72 &9.5&UX Ari&RS CVn\\
 FN3 &55662.965&2011 Apr 11 23:09:50& 03:26:06.13, +28:49:51.75 &0.31&0.26&0& 12 &14&UX Ari&RS CVn\\
 FN4 &55678.049&2011 Apr 27 01:10:25& 03:26:09.44, +28:47:41.58 &0.44&0.36&110& 32 &16&UX Ari&RS CVn\\
 FN5 &55219.221&2010 Jan 23 05:18:10& 03:37:02.86, +00:42:24.30 &0.5&0.43&170& 17 &24&HR1099&RS CVn\\
 FN6 &55244.054&2010 Feb 17 01:17:51& 03:36:28.02, +00:23:58.80 &0.43&0.39&90& 0.53 &7.1&HR1099&RS CVn\\
 FN7 &55503.647&2010 Nov 03 15:31:35& 03:39:24.90, +00:19:37.43 &1.1&0.73&40& 4.5 &5.1&HR1099&RS CVn\\
 FN8 &55625.878&2011 Mar 05 21:03:45& 03:36:24.15, +00:31:51.98 &0.31&0.21&12& 47 &29&HR1099&RS CVn\\
 FN9 &55510.015&2010 Nov 11 00:21:07& 11:40:43.67, -65:01:44.00 &0.44&0.35&15& 0.27 &10&GT Mus&RS CVn\\
 FN10 &55769.137&2011 Jul 27 03:16:43& 14:34:07.69, -60:33:54.43 &0.49&0.32&0& 58 &8.7&V841 Cen&RS CVn\\
 FN11 &55219.255&2010 Jan 23 06:06:55& 22:08:29.83, +45:38:19.77 &0.64&0.54&75& 22 &17&AR Lac&RS CVn\\
 FN12 &55376.647&2010 Jun 29 15:31:51& 22:06:42.49, +45:40:20.19 &0.75&0.56&130& 19 &6.7&AR Lac&RS CVn\\
 FN13 &55785.823&2011 Aug 12 19:45:20& 22:14:56.90, +46:15:13.01 &2.4&0.96&45& 0.71 &5.2&AR Lac&RS CVn\\
 FN14 &55101.059&2009 Sep 28 01:25:30& 23:11:50.76, +02:55:10.28 &0.53&0.49&15& 3.5 &5.7&SZ Psc&RS CVn\\
 FN15 &55063.937&2009 Aug 20 22:29:55& 23:58:33.24, +28:15:57.20 &1.1&0.97&45& 2.0 &19&II Peg&RS CVn\\
 FN16 &55291.166&2010 Apr 05 03:58:30& 23:54:58.71, +28:51:27.99 &0.74&0.64&20& 41 &13&II Peg&RS CVn\\
 FN17 &55433.679&2010 Aug 25 16:17:55& 23:55:01.76, +28:36:48.00 &1.1&0.7&153& 1.2 &9.9&II Peg&RS CVn\\
 FN18 &55434.315&2010 Aug 26 07:34:00& 23:57:52.31, +28:39:34.00 &1.2&9.5&0& 0.29 &9.6&II Peg&RS CVn\\
 FN19 &55561.057&2010 Dec 31 01:22:00& 03:07:57.26, +40:50:15.00 &0.46&0.33&10& 20 &11&Algol&Algol\\
 FN20 &55613.840&2011 Feb 21 20:10:00& 20:43:34.04, -32:13:03.96 &0.56&0.52&45& 69 &19&AT Mic&dMe\\
 FN21 &55574.278&2011 Jan 13 06:41:00& 23:31:33.68, +19:43:53.94 &0.35&0.32&40& 36 &15&EQ Peg&dMe\\
 FN22 &55628.897&2011 Mar 08 21:31:00& 07:43:46.70, +03:26:26.16 &0.32&0.29&160& 39 &45&YZ CMi&dMe\\
 FN23 &55446.767&2010 Sep 07 18:25:00& 10:43:41.64, -33:36:30.75 &0.57&0.45&150& 14 &17&TWA 7&YSO\\
\hline
\end{tabular} }
\end{center}
{\footnotesize 
\footnotemark[$*$] The time when the maximum luminosity was observed.\\
\footnotemark[$\dagger$] Observed count rate in the 2--10 keV band.
}
\label{FN}
\end{table*}
\end{landscape}

\clearpage

\begin{landscape}
\begin{table*}[h]
\caption{General properties of stars in our sample.}
\begin{center}
{\scriptsize
\begin{tabular}{lccccccccc}
\hline \hline
Object name & HD & Spectral type & Rotation velocity & Radius &
 {\it a} sin{\it i}\footnotemark[$*$]& inclination of orbit & $e$ &Distance & References \\
 & & & (km s$^{-1}$) & (R$_{\odot}$) & (R$_{\odot}$) & (degree) && (pc) & \\ \hline
VY Ari & 17433 & K3 V + K4 I\hspace{-.1em}V$^{\mathsection}$ & 10.2 & 1.90 &
		     8.2 & 57 &0.085 &44 & (1)(2)(3)(4) \\
UX Ari & 21242 & G5 V + K0 I\hspace{-.1em}V$^{\mathsection}$ & 41.5 & 5.78 & 5.9 (h) /
		     5.3 (c) & 59.2 & 0 &50.2 & (3)(4)(5)(6) \\
HR1099 & 22468 & G5 I\hspace{-.1em}V + K1 I\hspace{-.1em}V$^{\mathsection}$ & 59.0
	     & 3.30 & 1.9 (h) / 2.4 (c) & 38 & 0 & 29 & (4)(7)(8)(9)(10)\\
GT Mus & 101379 & (A0V+A2V) + (G5III+G8III$^{\mathsection}$) & -- & 33.0 &
		     15 & 10 & 0.032 & 172 & (4)(11)(12)(13) \\
V841 Cen\footnotemark[$\ddagger$] & 127535 & K1
	 I\hspace{-.1em}V$^{\mathsection}$& -- & 3.8 & 4.1 & -- & 0 & 63 & (14)(15)(16) \\ 
AR Lac & 210334 & G2 I\hspace{-.1em}V + K0 I\hspace{-.1em}V$^{\mathsection}$ & 73.7
	     & 2.68 & 4.6 (h) / 4.5 (c) & 89.4 & 0 & 42 & (3)(4)(17)(18)(19) \\
SZ Psc & 219113 & F5 I\hspace{-.1em}V + K1 I\hspace{-.1em}V$^{\mathsection}$ &
	     80.2 & 6.0 & 8.7 (h) / 6.4 (c) & 69.8 & 0 & 88 & (3)(4)(20)(21) \\
II Peg & 224085 & K2 I\hspace{-.1em}V$^{\mathsection}$ + M0-3 V & -- & 2.21
		 & 4.9 & 60 &0& 42 & (4)(22)(23)(24)(25) \\
Algol & 19356 & B8 V + K2 I\hspace{-.1em}V$^{\mathsection}$ & -- & 3.4 &
		     14 & 81.4 & 0 & 28.5 & (26)(27)(28)(29) \\
AT Mic & 196982 & M4.5 V + M4.5 V$^{\mathsection}$ & 24.6 & 0.38 & 5980 & -- & -- &10.2 & (30)(31)(32)(33) \\
EQ Peg & -- & M3.5 V + M5 V$^{\mathsection}$ & 88.5 & 0.35 & 5590 & 30 & -- &6.5 & (34)(35)(36)(37)(38) \\
YZ CMi & -- & M4.5 V$^{\mathsection}$ & 5.3 & 0.29 & -- & -- & -- & 5.9 & (31)(34)(39) \\
TWA-7 & -- & M2 V$^{\mathsection}$ & 19.2 & 1.89 & -- & -- & -- & 27 & (40)(41)\\
 \hline
\end{tabular} }
\end{center}
{\footnotesize 
\footnotemark[$*$]The {\it a} and {\it i} in {\it a} sin{\it i} show the
semi-major axes and inclination angle, respectively. The (h) and (c) show
the semi-major axes for the orbit of the hot and cool components,
respectively.\\
\footnotemark[$\mathsection$] The stellar component with the symbol $\S$ has
 the largest radius in the multiple-star system. As for the values for the
 radius and the rotation velocity in this table, the components with
 this symbol are  used.\\
\footnotemark[$\ddagger$] V841 Cen is a single-lined RS CVn binary.
\newline
(1) \cite{Alekseev+01} (2) \cite{Bopp+89} (3) \cite{Glebocki+95}
 (4) \cite{ESA97} (5) \cite{Carlos+71} (6) \cite{Duemmler+01}
 (7) \cite{Donati+97} (8) \cite{Donati99} (9) \cite{Fekel83} (10)
 \cite{Lanza+06} (11) \cite{Murdoch+95} (12) \cite{Randich+93}
 (13) \cite{Stawikowski+94} (14) \cite{Ozeren+99} (15)
 \cite{Strassmeier+94} (16) \cite{Collier82} (17) \cite{Chambliss76}
 (18) \cite{Zboril+05} (19) \cite{Sanford51} (20) \cite{Eaton+07} (21)
 \cite{Jakate+76} (22) \cite{Berdyugina+98} (23) \cite{Marino+99} (24)
 \cite{ONeal+01} (25) \cite{Vogt81} (26) \cite{Richards+03} (27)
 \cite{Sarna93} (28) \cite{Singh+95} (29) \cite{Oord+89} (30)
 \cite{Lim+87} (31) \cite{Mitra07} (32) \cite{Mitra+05} (33)
 \cite{Wilson78} (34) \cite{Morin+08} (35) \cite{Pallavicini+90a} (36)
 \cite{Pettersen+84} (37) \cite{Pizzolato+03} (38) \cite{Hopmann58} (39) \cite{Reiners+09} 
 (40) \cite{Mamajek05} (41) \cite{Yang+08} }
\label{prop}
\end{table*}
\end{landscape}

\clearpage
{\scriptsize
\begin{landscape}
\begin{longtable}{l ccccc ccccc}
\caption{Best-fit parameters in the fitting, and  derived flare
 parameters.}
 \label{para}
\\
\hline
\hline
 & $kT$ & $EM$ & Flux$^{*}$ & $L_{\rm
 X(2-20)}^{\dagger}$ & $\chi^{2}_{\nu}$ (d.o.f) & ${\tau}_{d}^{\ddagger}$ & $E_{\rm tot}^{\S}$ & \multicolumn{2}{c}{$l^{\|}$} & $B^{\#}$ \\ 
\cline{9-10}
 & (keV) & ($10^{54}$& ($10^{-10} {\rm
	     ergs}$ &  ($10^{31}$& & (ks)& ($10^{35}$& (R$_{\odot}$)&(binary separation)& (G)\\
 &  & ${\rm cm}^{-3}$) & ${\rm cm}^{-2}\hspace{1ex} {\rm s}^{-1}$)
	     & $ {\rm ergs\hspace{1ex} s}^{-1}$) & & &
			      ergs) &  &  &  \\ \hline
\endhead
\hline
\endfoot
\hline
\multicolumn{11}{l}{\hbox to 0pt{\parbox{180mm}{\footnotesize Since statistics are too limited, FN13 is not
 fitted. Abundances are fixed to the cosmic value, and the absorbing
 columns are fixed to zero.  Errors and lower limits refer to 90 \%
 confidence intervals. The parameters $kT$, $EM$, Flux, $L_{\rm X}$ of FN23 (TWA-7) are
 from \citet{Uzawa+11}.\\ 
${*}$ Flux in the 2--20 keV band.\\
$\dagger$ Absorption-corrected $L_{\rm X}$ in the 2--20
 keV band. Distances are assumed to be the corresponding values in table
 1. \\
$\ddagger$ {\it e}-folding time derived with light-curve
 fitting with a burst model (linear rise and exponential decay).\\
$\S$ Total released energy derived by multiplying $L_{\rm
 X}$ by $\tau_{d}$. \\
$\|$ Loop length of flare. \\
$\#$ Magnetic-field strength.\\}}}
\endlastfoot 
FN1 & 5 & 1 & 2 & 5 & 0.42 (4) & 26 & 10 & 4 &
 0.4 & 50 \\ & (2--40) & (0.5--2) & (1--3) & (3--7) & &
 (11--50) & (8--20) & (0.1--20) & (0.01--2) & (10--3000) \\
 FN2 &
 4 & 24 & 8 & 20 & 1.7 (7) & 53 & 100 & 30 & 2 & 20 \\ &
 (2--30) & (15--36) & (4--10) & (10--30) & & (30--81) & (60--200) &
 (1--100) & (0.1--8) & (8--600) \\
 FN3 & ${\geq}$5 &
 60 & 40 & 100 & 0.63(6) & -- & -- & $\leq$50 &
 $\leq$4 & $\geq$20 \\ & & (40--80) & (30--50) &
 (90--200) & & -- & -- & & & \\
 FN4 & 3 & 100 & 40 & 100 & 0.60 (2) &
 24 & 300 & 100 & 10 & 10 \\ & (2--7) & (70--200) &
 (8--50) & (30--200) & & (10--42) & (60--400) & (20--500) &
(2--40) & (3--50) \\ 
 FN5 & 5 & 30 & 30 & 30 & 0.42 (5) &
 6 & 20 & 20 & 3 & 30 \\ & (3--10) & (20--40) & (10--40)
 & (10--40) & & (4--8) & (5--30) & (5--70) & (0.8--10) &
 (10--100) \\
 FN6 & ${\geq}$4 & 3 & 6 & 6 & 0.5 (7) & 67 & 40 &
 $\leq$10 & $\leq$2 & $\geq$30 \\ &  & (2--4) &
 (5--7) & (5--7) & & (45--93) & (30-50) & & & \\
 FN7 & ${\geq}$4 &
 4 & 8 & 8 & 0.76 (3) & -- & -- & $\leq$10 &
 $\leq$ 10 & $\geq$30 \\ & & (3--6) & (6--11) &
 (6--11) & & -- & -- & & & \\
 FN8 & 7 & 30 & 50 & 50 & 0.33 (3) & 14 &
 70 & 20 & 2 & 50 \\ & (3--30) & (20--50) & (2--60) &
 (2--60) & & (11--18) & (2--80) & (2--70) & (0.2--10) &
 (10--500) \\
 FN9 & 8 & 160 & 8 & 270 & 1.8 (7) & 130 & 3500 & 50 &
 0.6 & 40 \\ & (4--20) & (130--210) & (2--9) & (90--320) & &
 (89--190) & (1200--4200) & (7--600) & (0.1--7) & (5--300) \\
 FN10 & ${\geq}$4 & 40 & 20 & 100 & 0.19 (3) & 7 & 70 & $\leq$60 &
 -- & $\geq$20 \\ & & (20--60) & (10--30) & (60--200) & & (0.1--15) &
 (3--200) & & & \\
 FN11 & ${\geq}$6 & 24 & 30 & 60 & 0.84 (7) & 3 &
 16 & $\leq$20 & $\leq$2 & $\geq$50 \\ & & (18--31) &
 (20--35) & (40--70) & & (1--4) & (8--23) & & & \\
 FN12 &
 ${\geq}$2 & 30 & 30 & 70 & 1.0 (4) & 7 & 50 & $\leq$200 &
 $\leq$20 & $\geq$4 \\ & & (10--40) & (20--70) &
 (30--140) & & (4--10) & (20--100) & & & \\
 FN14 & ${\geq}$2 & 50 &
 6 & 50 & 1.2 (8) & 72 & 400 & $\leq$300 & $\leq$20 &
 $\geq$5 \\ & & (30--130) & (4--7) & (40--70) & & (49--100) &
 (300--500) & & & \\
 FN15 & ${\geq}$8 & 200 & 230 & 500 & 1.6 (4) & 19
 & 900 & $\leq$50 & $\leq$8 & $\geq$50 \\ & & (100--300) & (160--420)
 & (300--900) & & (15--24) & (700--2000) & & & \\
 FN16 & 3 & 40 & 10 &
 25 & 1.1 (5) & 6 & 14 & 90 & 20 & 10 \\ & (2--7) & (20--60) &
 (5--20) & (11--32) & & (4--8) & (6--20) & (10--300) & (2--50) &
 (4--70) \\
 FN17 & ${\geq}$7 & 17 & 20 & 40 & 0.3 (6) & 12 & 50 &
 $\leq$10 & $\leq$2 & $\geq$60 \\ & & (10--23) & (10--40) & (30--80) &
 & (5--43) & (30--160) & & & \\
 FN18 & 4 & 8 & 4 & 8 & 1.3 (3)
 & 41 & 30 & 20 & 3 & 30 \\ & (2--10) & (4--12) & (0.2--5) &
 (0.4--10) & & (19--97) & (2--60) & (3--70) & (0.5--10) & (8--200)
 \\
 FN19 & -- & 20 & 20 & 30 & 0.25 (4) & 5 & 10 & -- & -- & -- \\
 & -- & (10--30) & (10--40) & (20--60) & & (2--9) & (6--30) & & & \\
 FN20 & ${\geq}$6 & 2.5 & 50 & 6 & 0.92 (5) & 6 & 3 & $\leq$5 & -- &
 $\geq$70 \\
 & & (1.8--3.2) & (30--60) & (4--8) & & (4--9) & (2--5)
 & & & \\
 FN21 & ${\geq}$4 & 1.0 & 40 & 2 & 0.54 (4) & 4 & 0.9 &
 $\leq$6 & $\leq$0.0005 & $\geq$40 \\
 & & (0.6--1.5) & (30--60) & (1--3)
 & & (2--8) & (0.4--1.7) & & & \\
 FN22 & 6 & 2 & 70 & 2.8 & 1.7
 (11) & ${\leq}$5 & ${\leq}$2 & 4 & -- & 80 \\
 & (3--15) &
 (1--3) & (30--80) & (1.1--3.3) & & -- & -- & (0.7--20) & &
 (20--400) \\ 
FN23 & 6 & 20 & 20 & 30 & 0.16 (3) & 6 & 20 & 20 & --
 & 50 \\ 
& (3--30) & (10--40) & (10--30) & (20--40) & & (3--14) &
 (10--40) & (1--80) & & (10--700) \\ 
\end{longtable}
\end{landscape}
}
\clearpage

\begin{figure*}[htbp]
\begin{minipage}{0.5\hsize}
\begin{center}
\includegraphics[width=75mm]{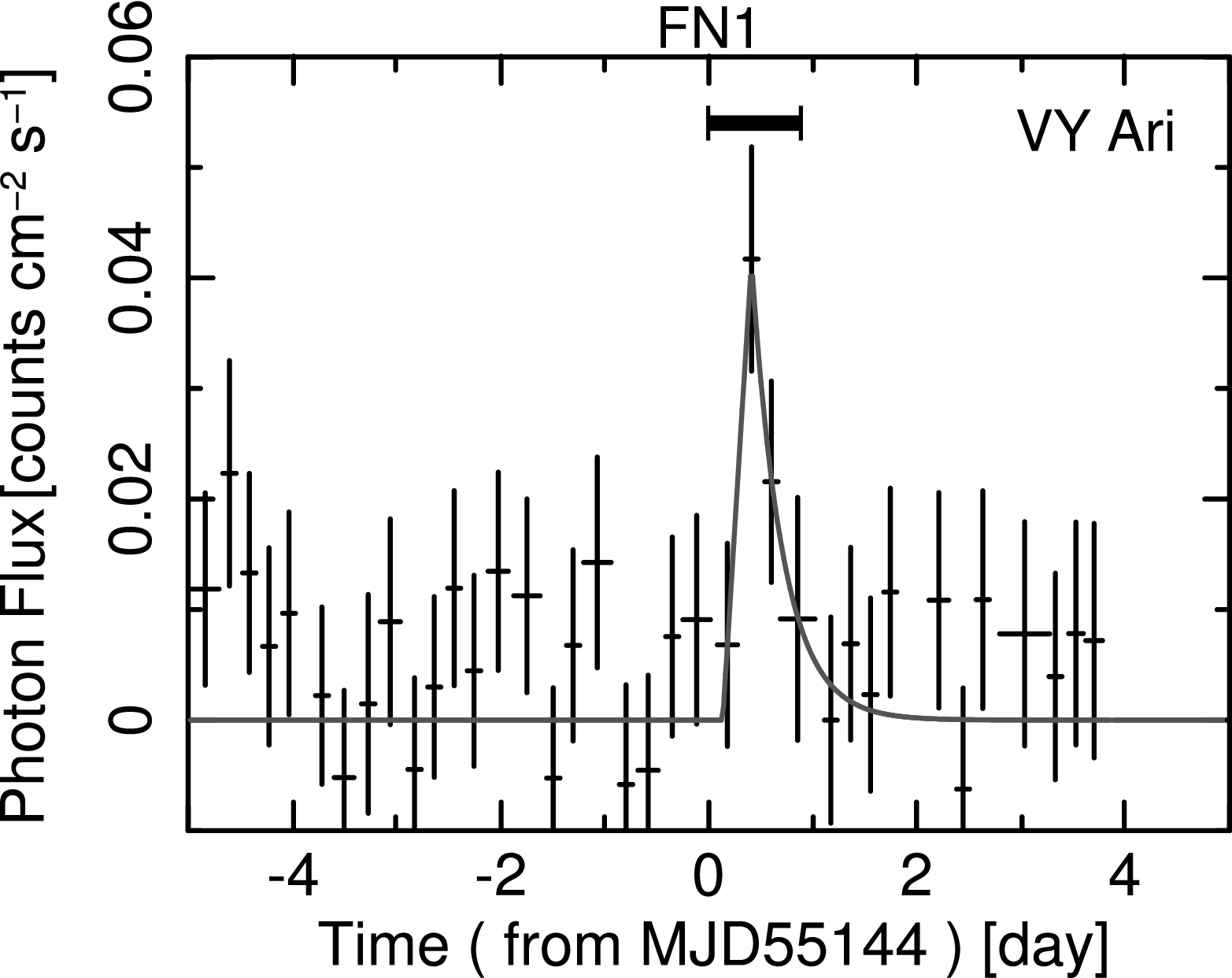}    
\end{center}
\end{minipage}
\begin{minipage}{0.5\hsize}
\begin{center}
\includegraphics[width=75mm]{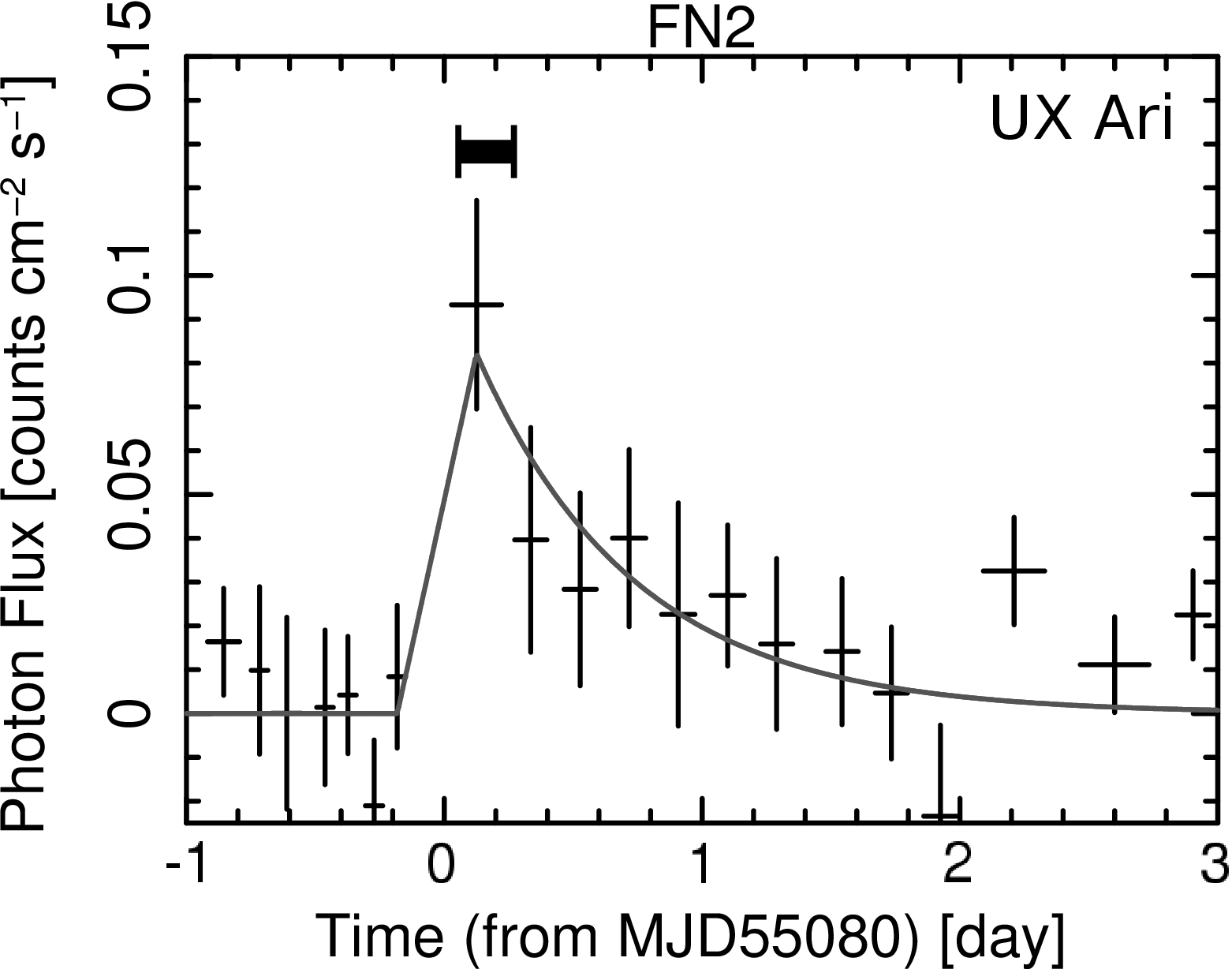}    
\end{center}
\end{minipage}
\begin{minipage}{0.5\hsize}
\begin{center}
\includegraphics[width=75mm]{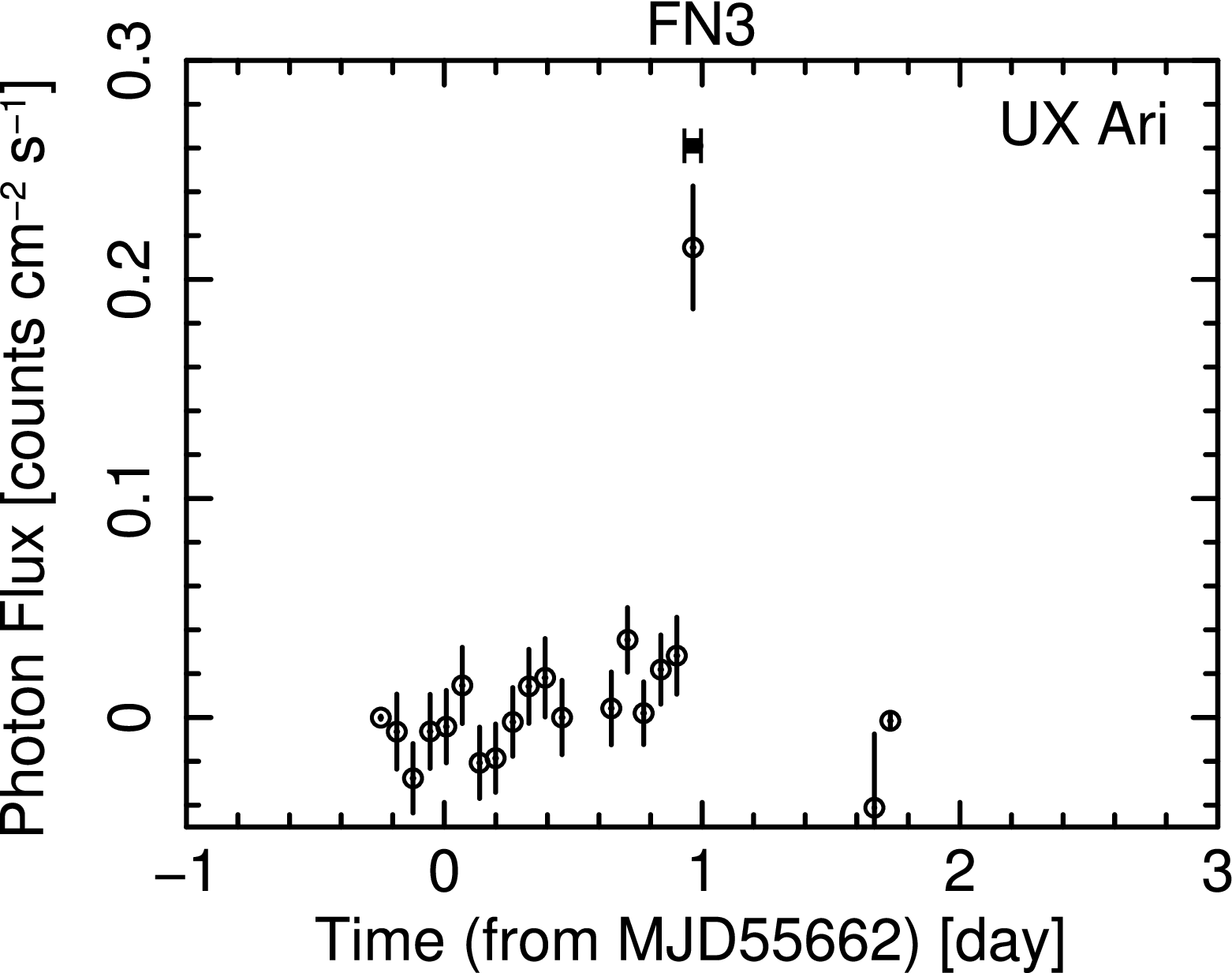}    
\end{center}
\end{minipage}
\begin{minipage}{0.5\hsize}
\begin{center}
\includegraphics[width=75mm]{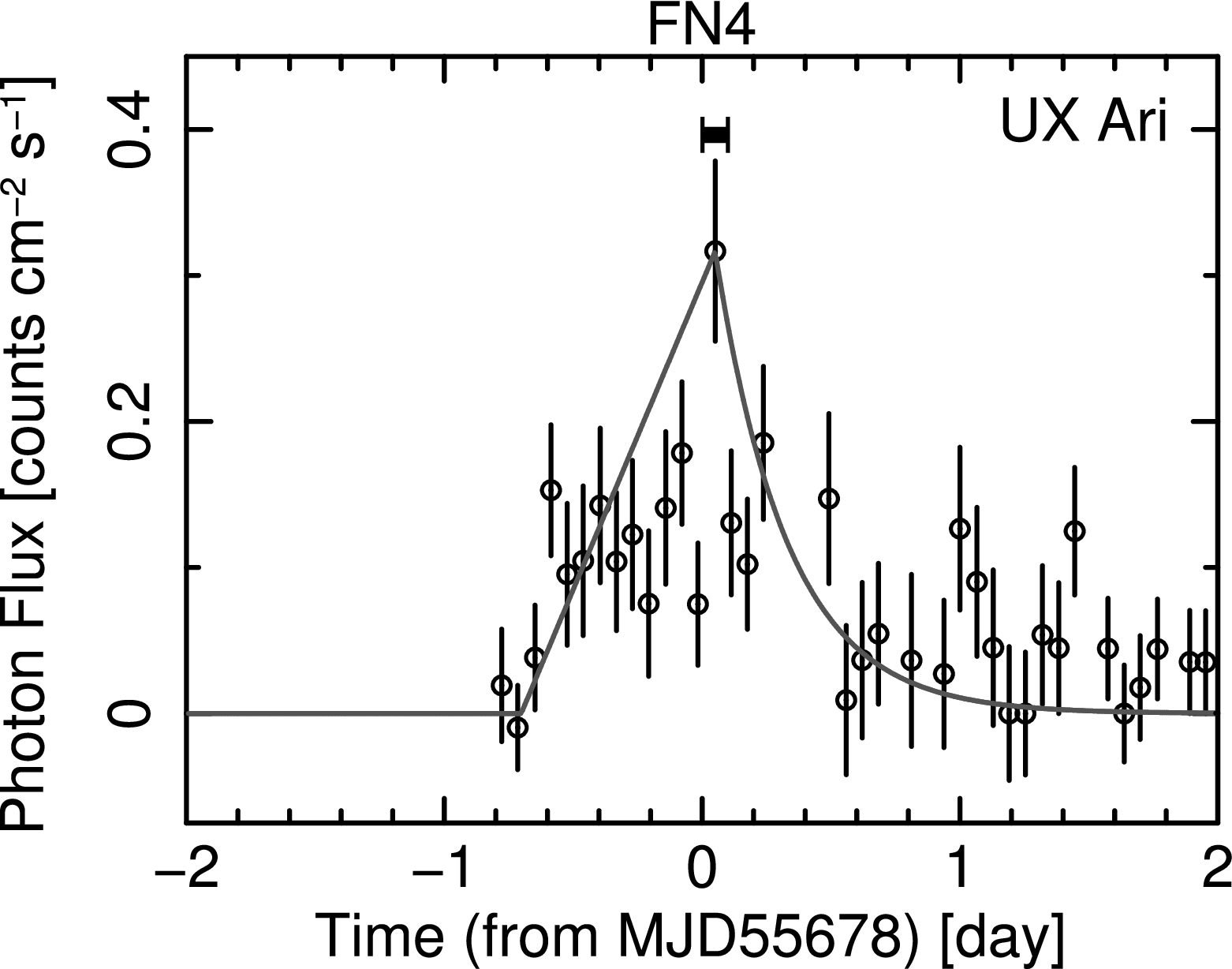}    
\end{center}
\end{minipage}
\begin{minipage}{0.5\hsize}
\begin{center}
\includegraphics[width=75mm]{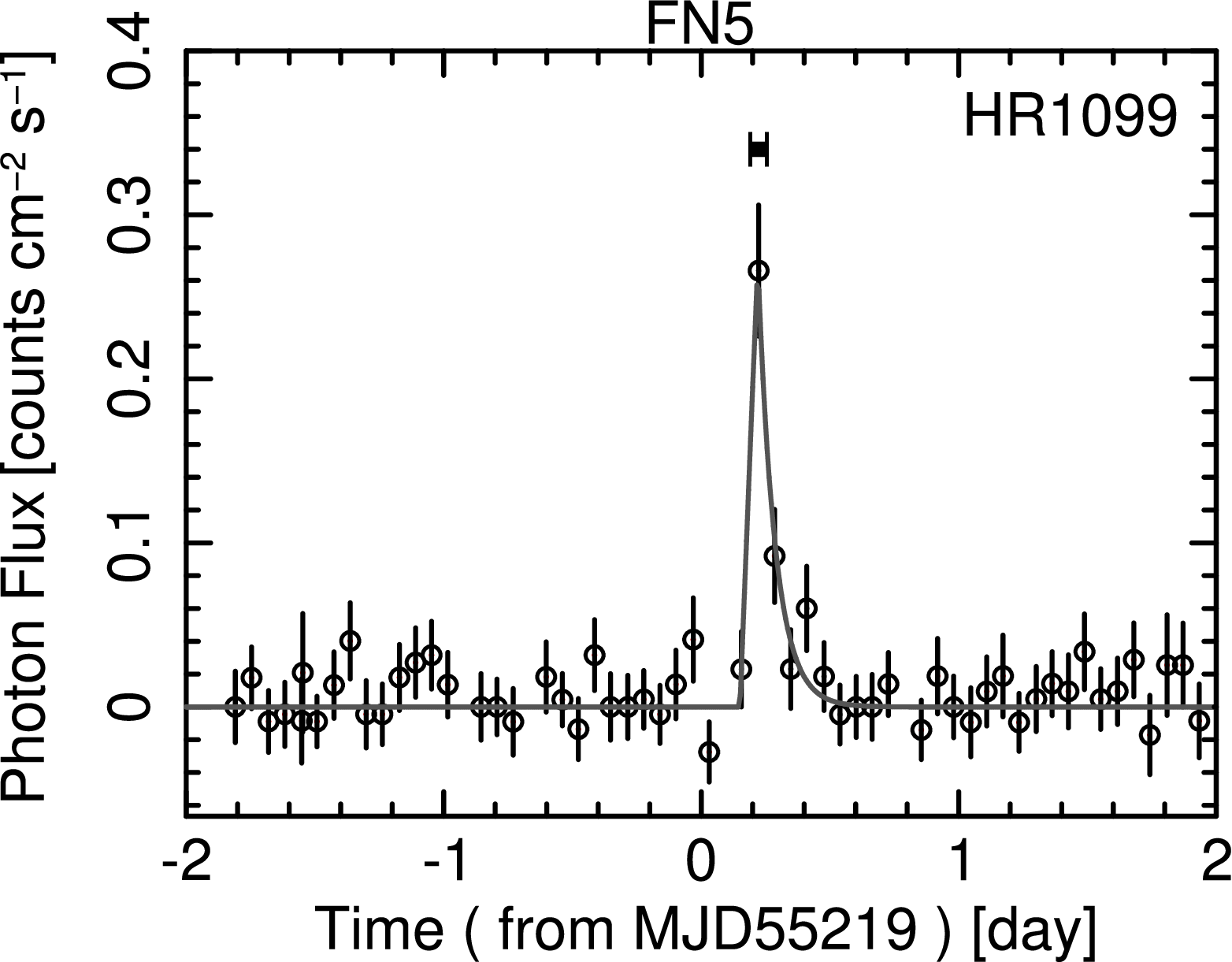}    
\end{center}
\end{minipage}
\begin{minipage}{0.5\hsize}
\begin{center}
\includegraphics[width=75mm]{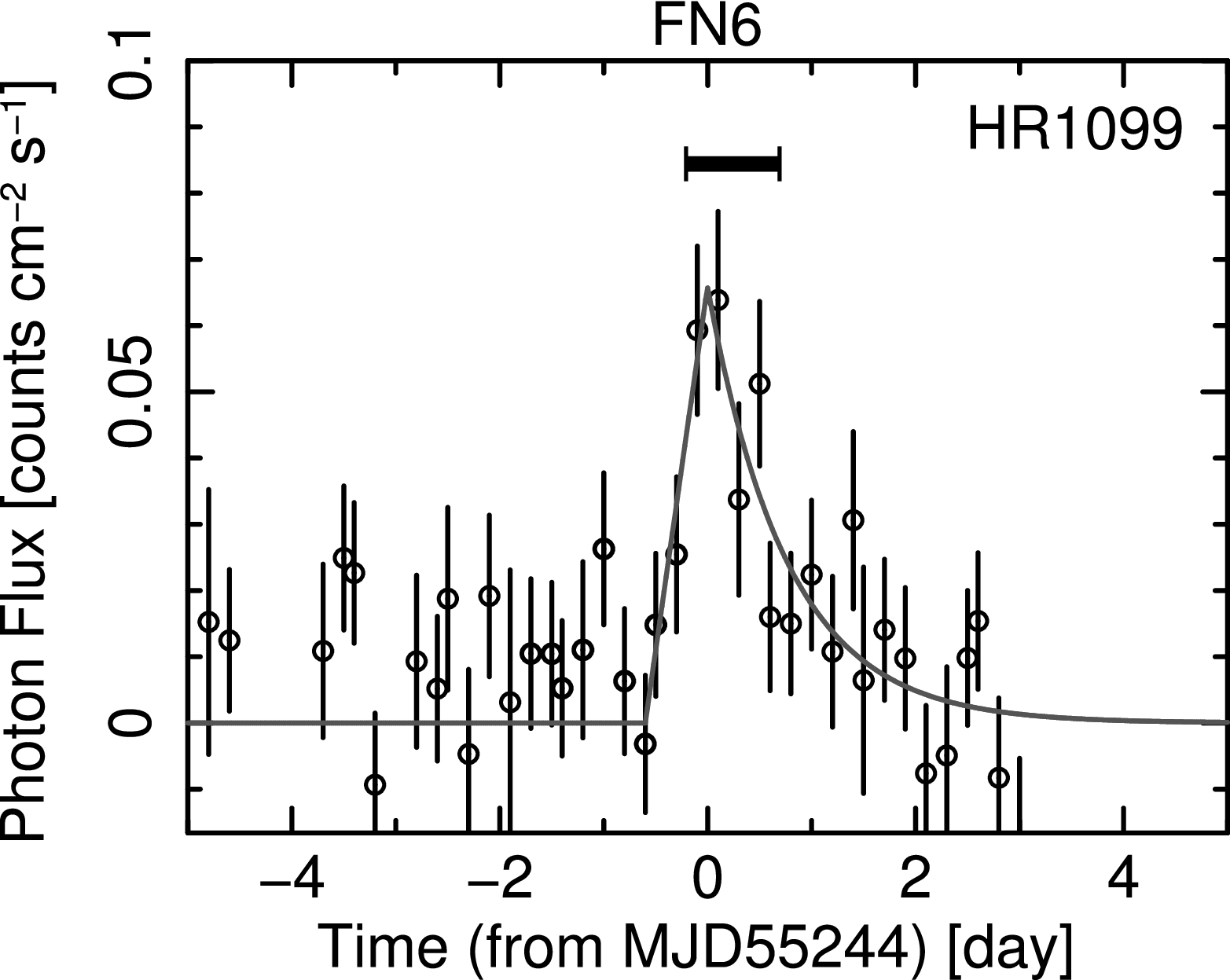}    
\end{center}
\end{minipage}
\end{figure*}
\clearpage
\begin{figure*}[htbp]
\begin{minipage}{0.5\hsize}
\begin{center}
\includegraphics[width=75mm]{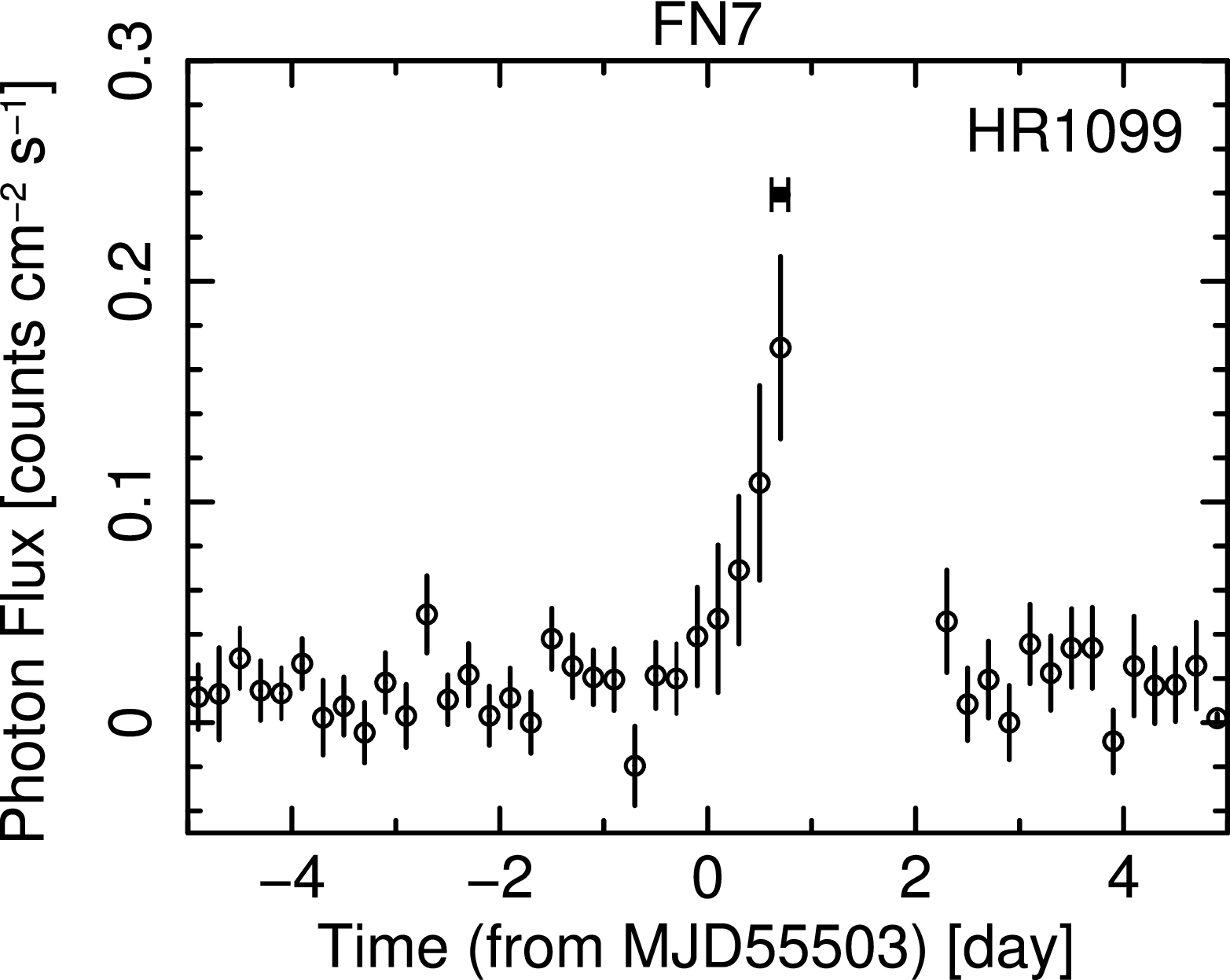}    
\end{center}
\end{minipage}
\begin{minipage}{0.5\hsize}
\begin{center}
\includegraphics[width=75mm]{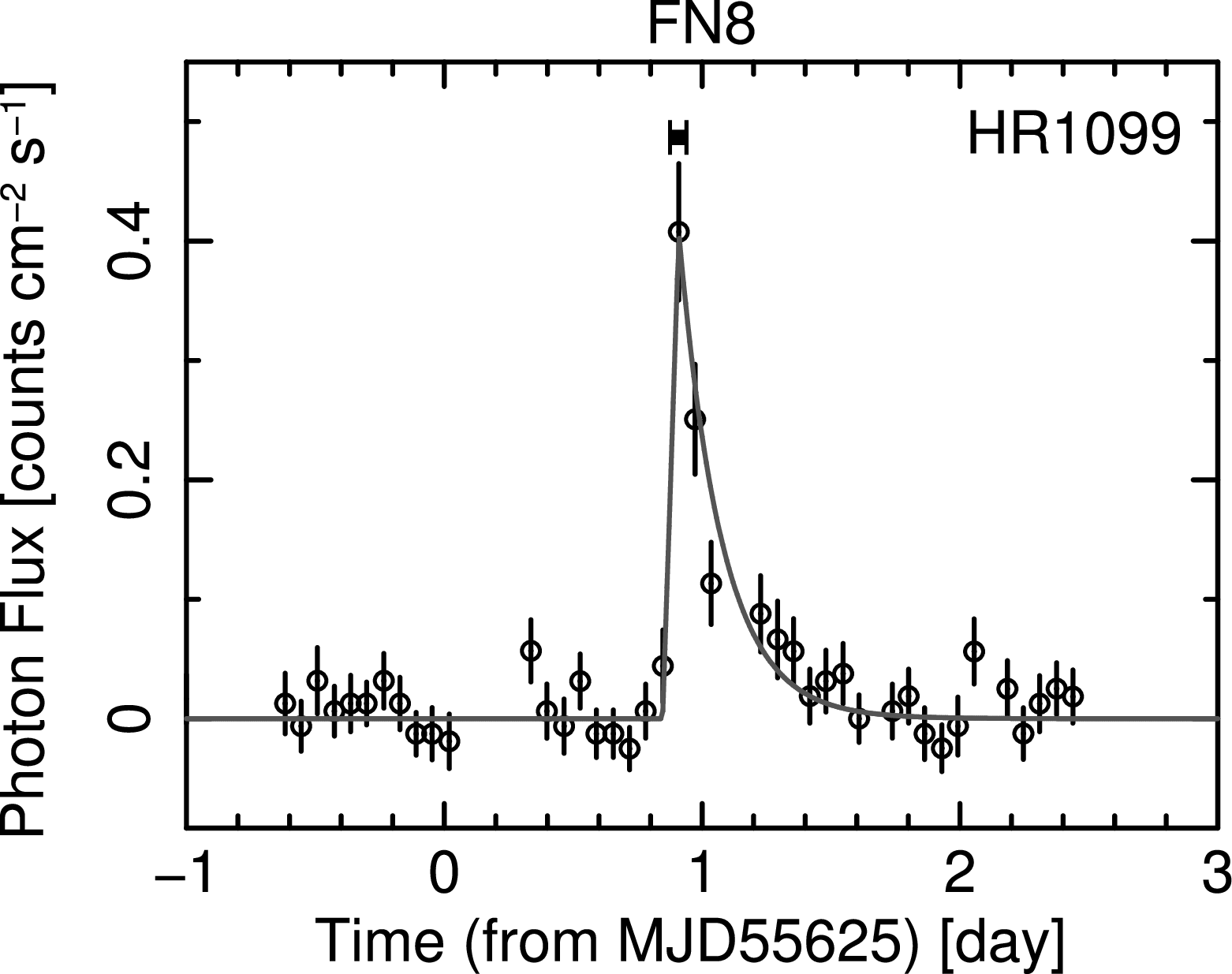}    
\end{center}
\end{minipage}
\begin{minipage}{0.5\hsize}
\begin{center}
\includegraphics[width=75mm]{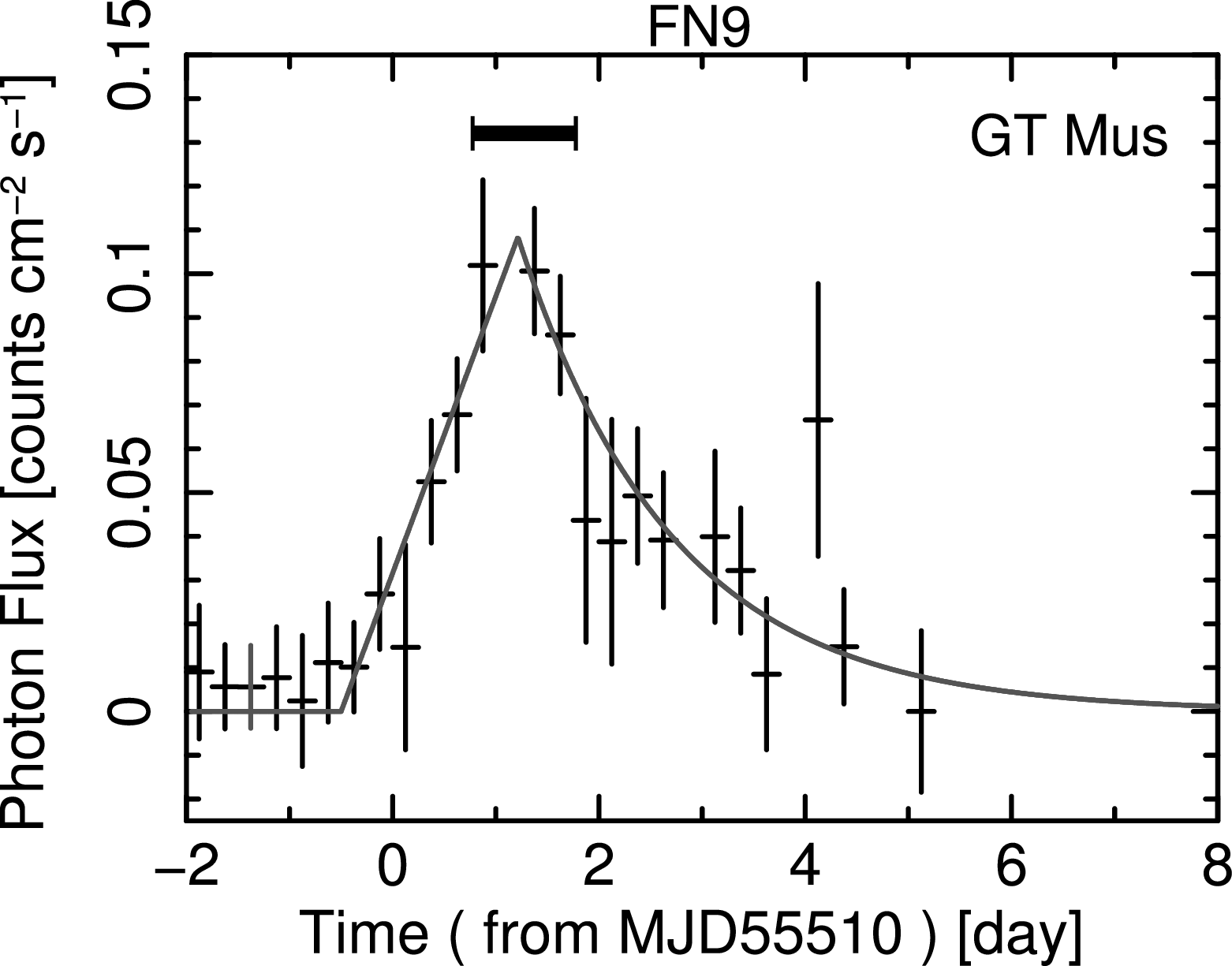}    
\end{center}
\end{minipage}
\begin{minipage}{0.5\hsize}
\begin{center}
\includegraphics[width=75mm]{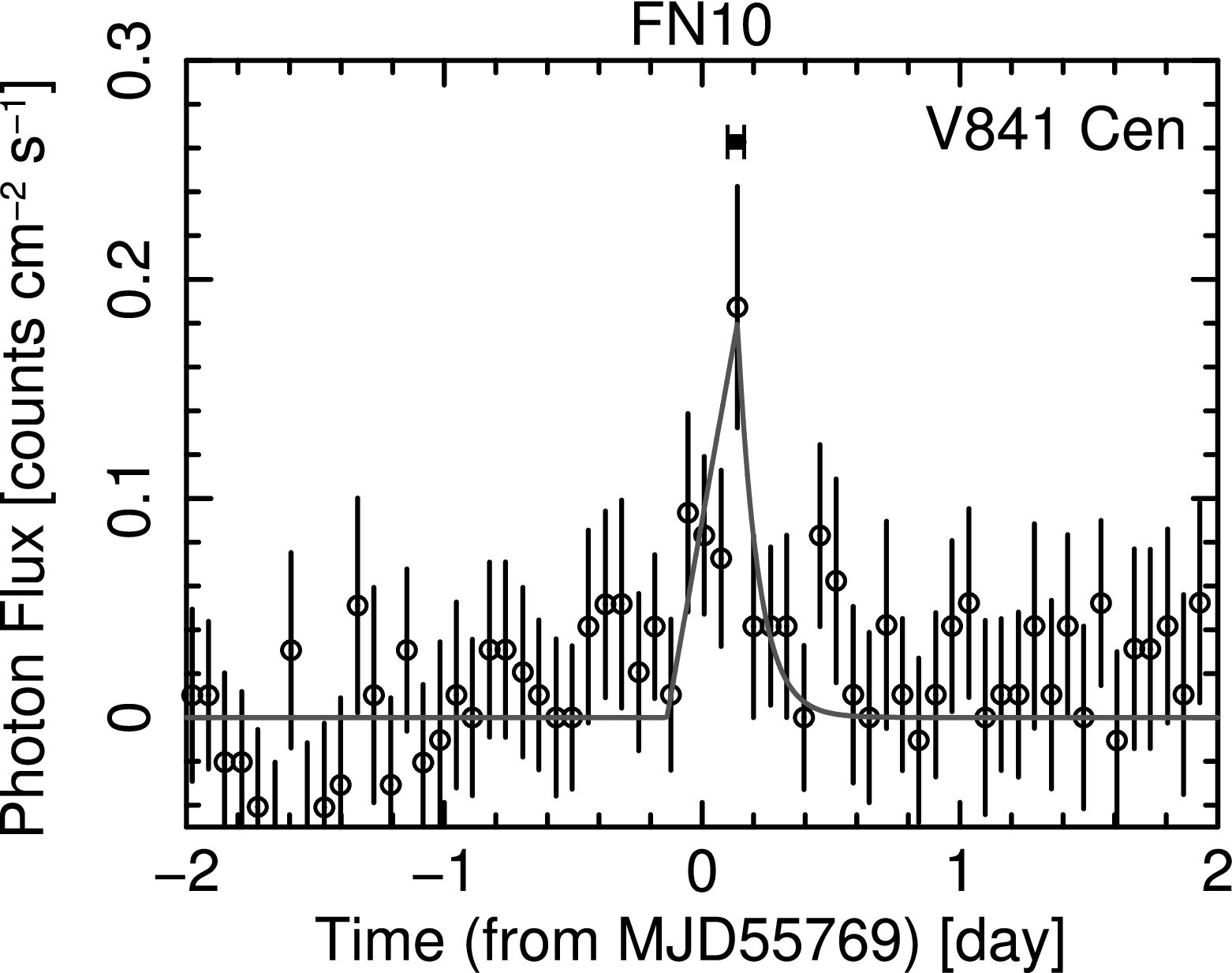}    
\end{center}
\end{minipage}
\begin{minipage}{0.5\hsize}
\begin{center}
\includegraphics[width=75mm]{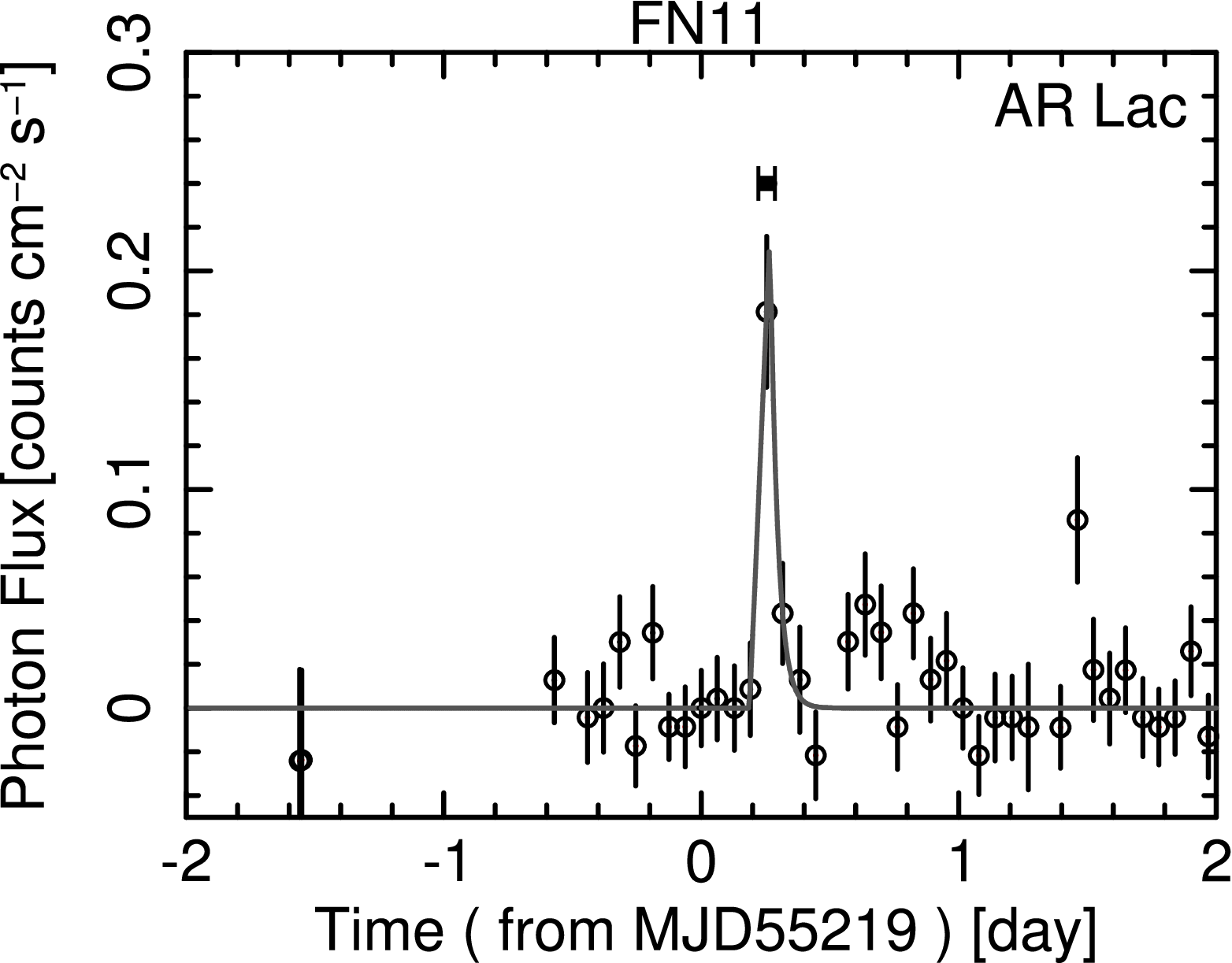}    
\end{center}
\end{minipage}
\begin{minipage}{0.5\hsize}
\begin{center}
\includegraphics[width=75mm]{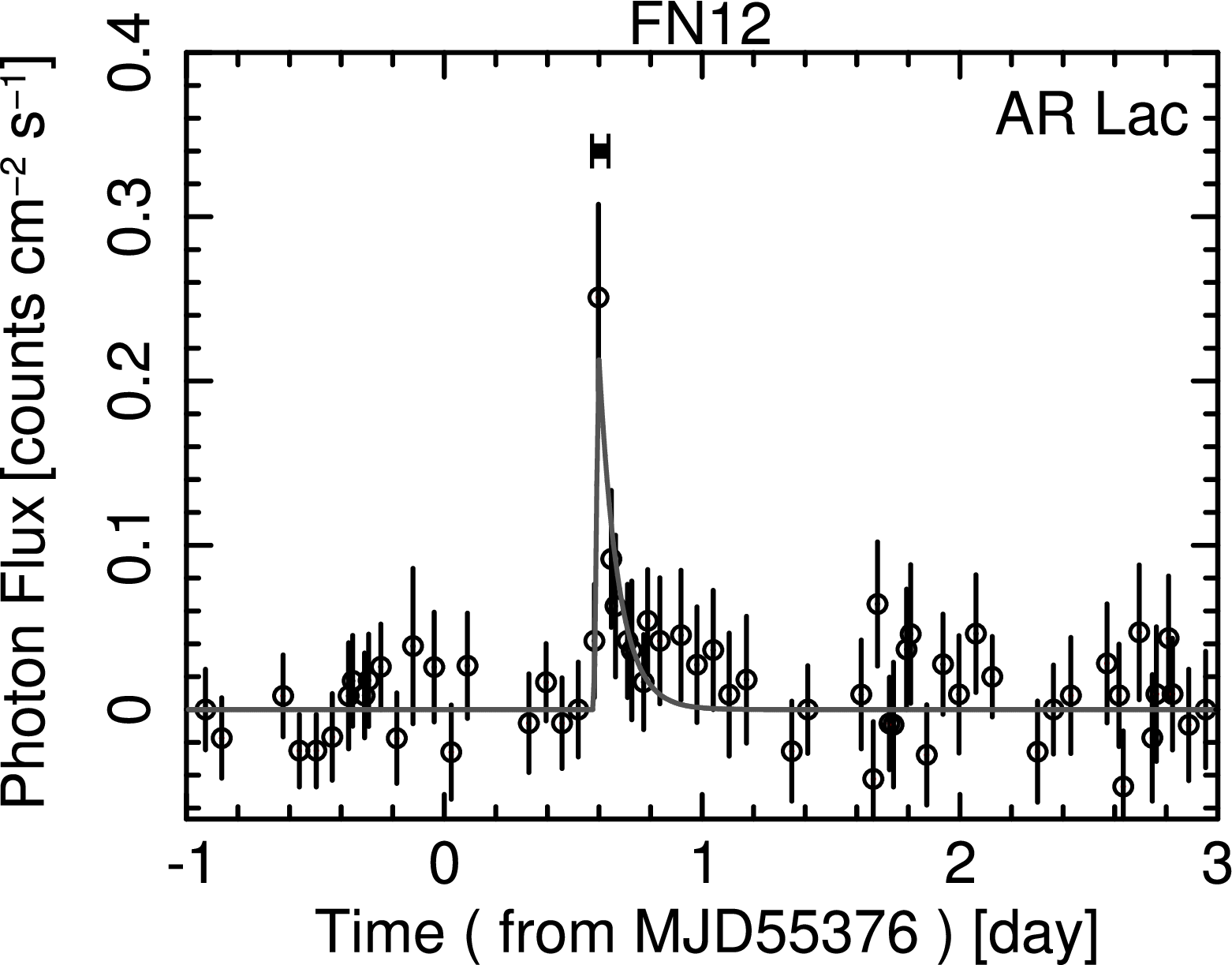}    
\end{center}
\end{minipage}
\end{figure*}
\clearpage
\begin{figure*}[htbp]
\begin{minipage}{0.5\hsize}
\begin{center}
\includegraphics[width=75mm]{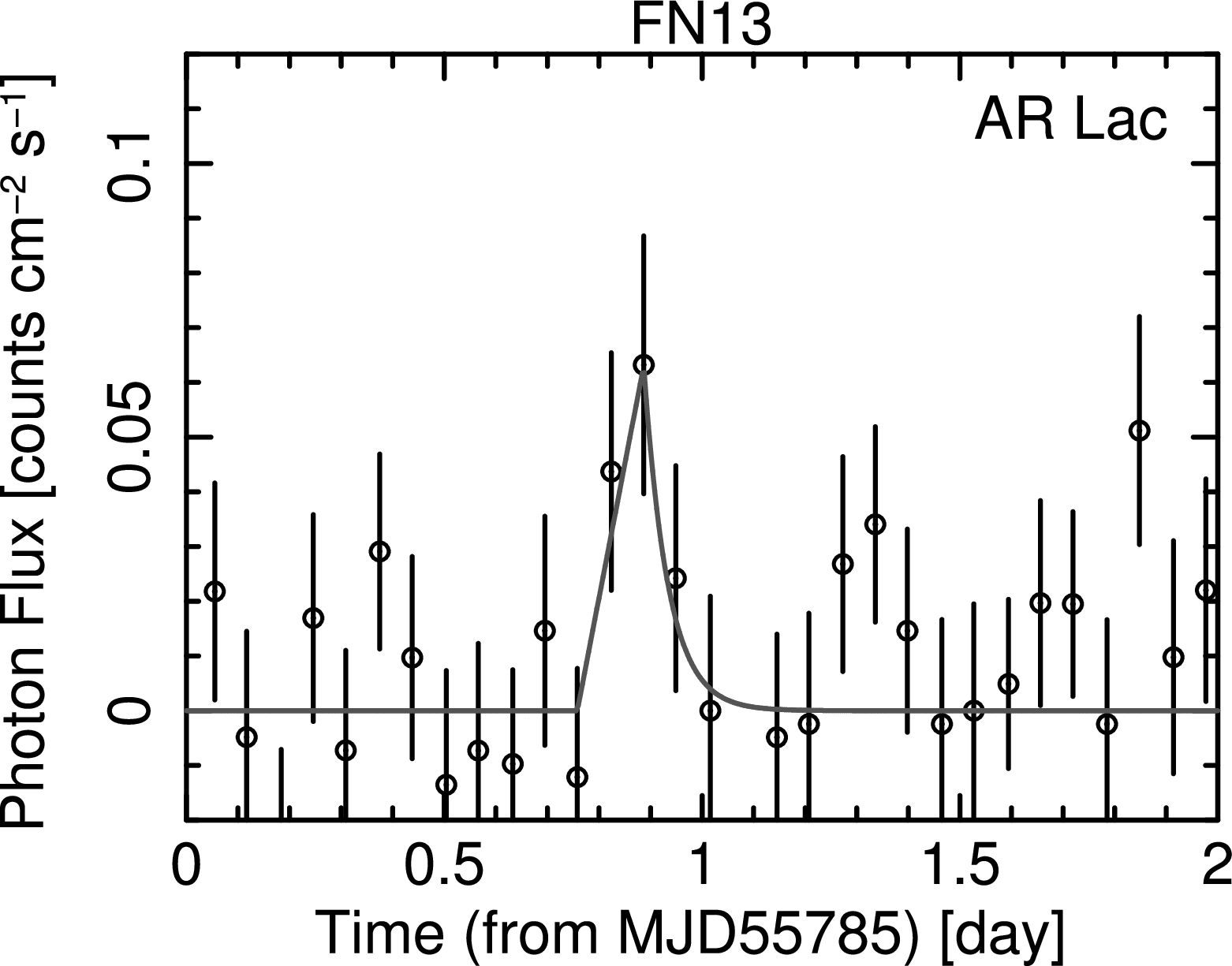}    
\end{center}
\end{minipage}
\begin{minipage}{0.5\hsize}
\begin{center}
\includegraphics[width=75mm]{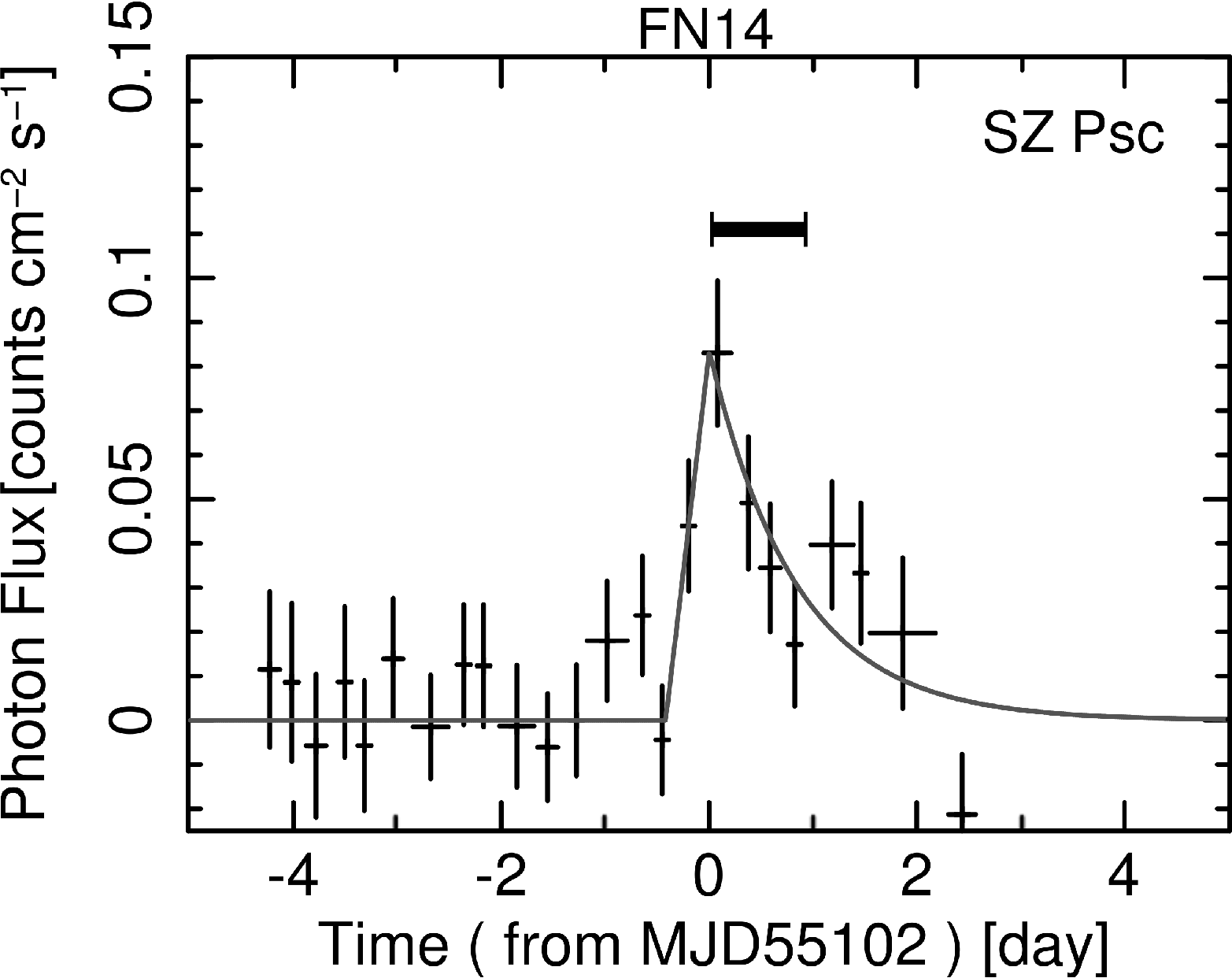}    
\end{center}
\end{minipage}
\begin{minipage}{0.5\hsize}
\begin{center}
\includegraphics[width=75mm]{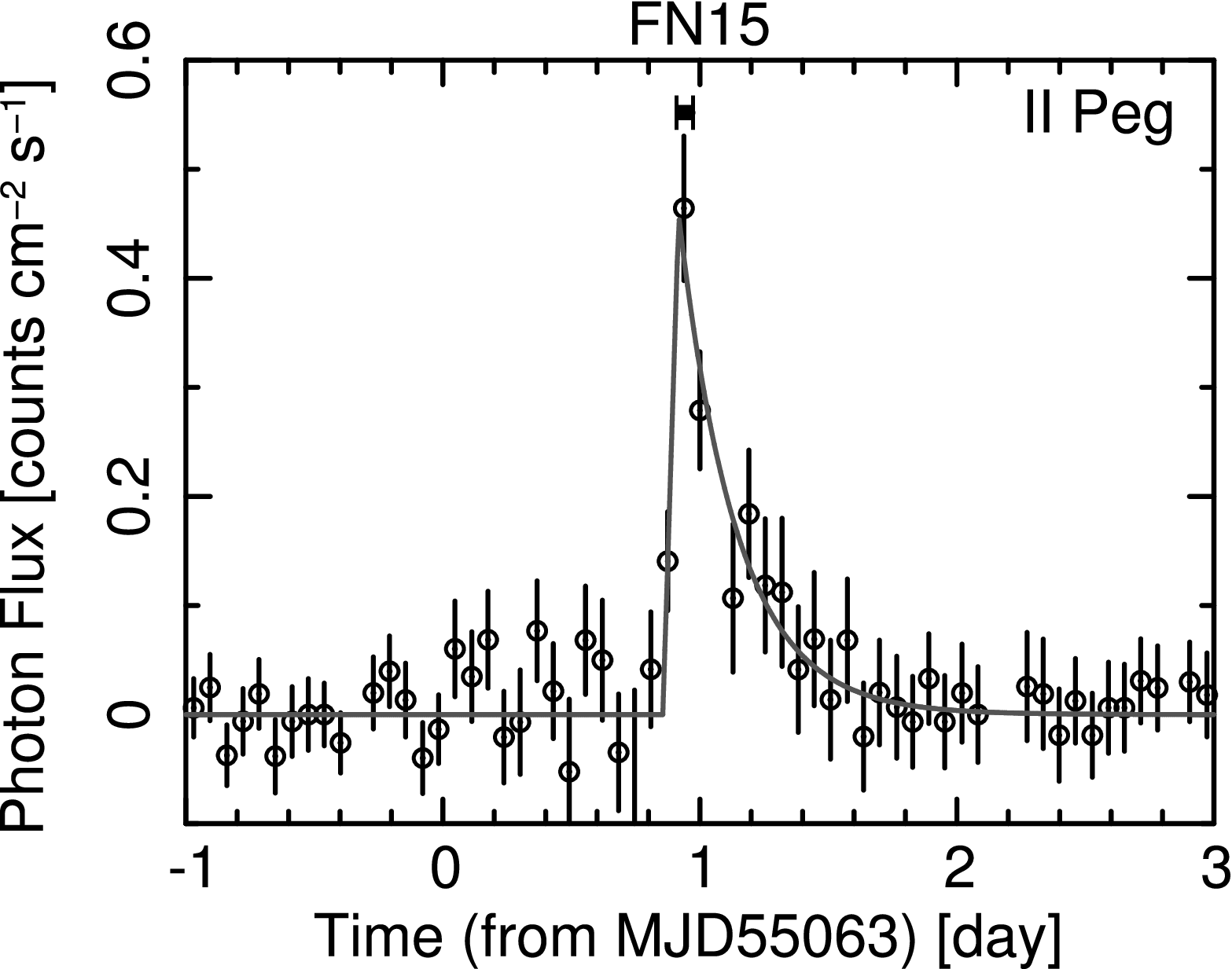}    
\end{center}
\end{minipage}
\begin{minipage}{0.5\hsize}
\begin{center}
\includegraphics[width=75mm]{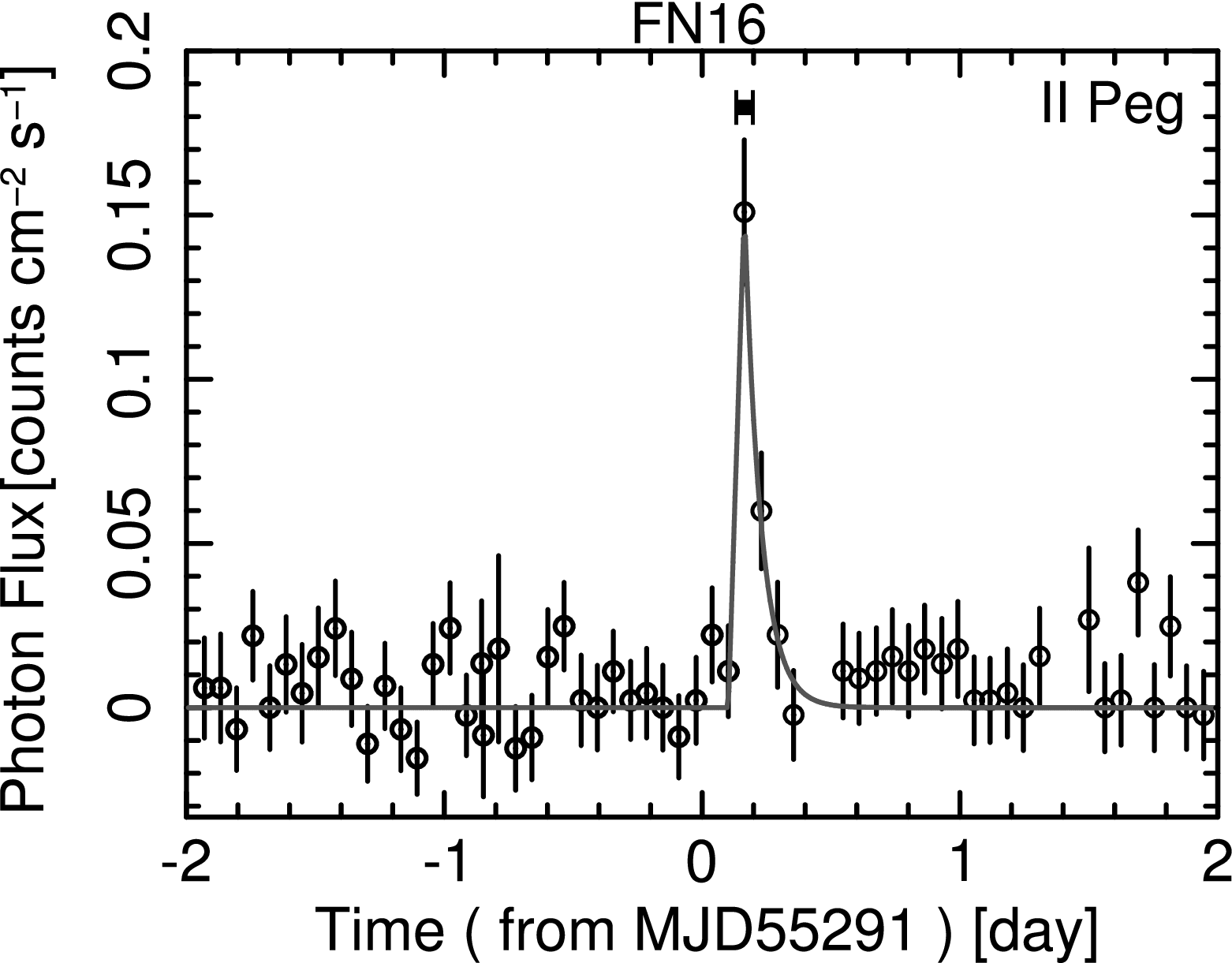}    
\end{center}
\end{minipage}
\begin{minipage}{0.5\hsize}
\begin{center}
\includegraphics[width=75mm]{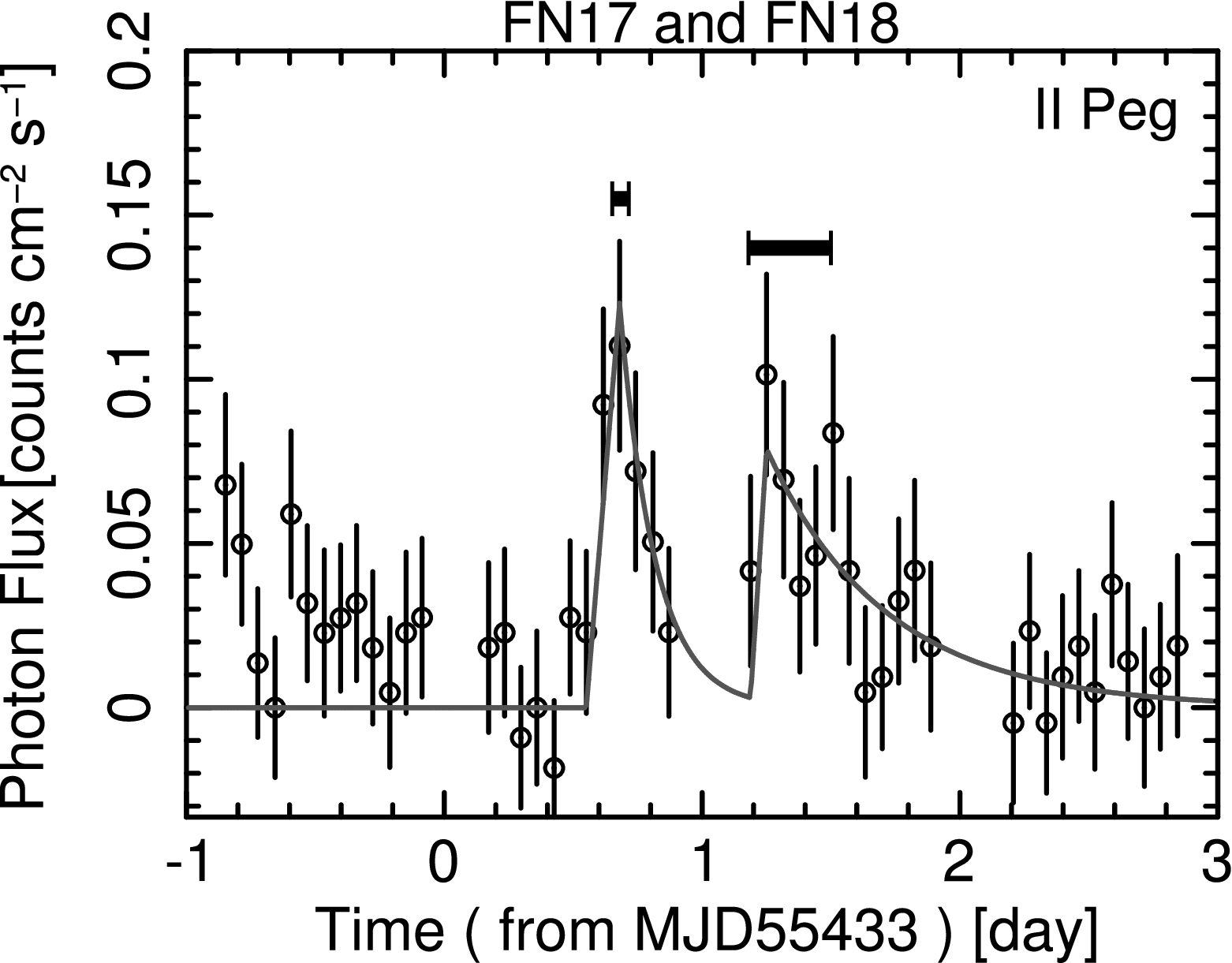}    
\end{center}
\end{minipage}
\begin{minipage}{0.5\hsize}
\begin{center}
\includegraphics[width=75mm]{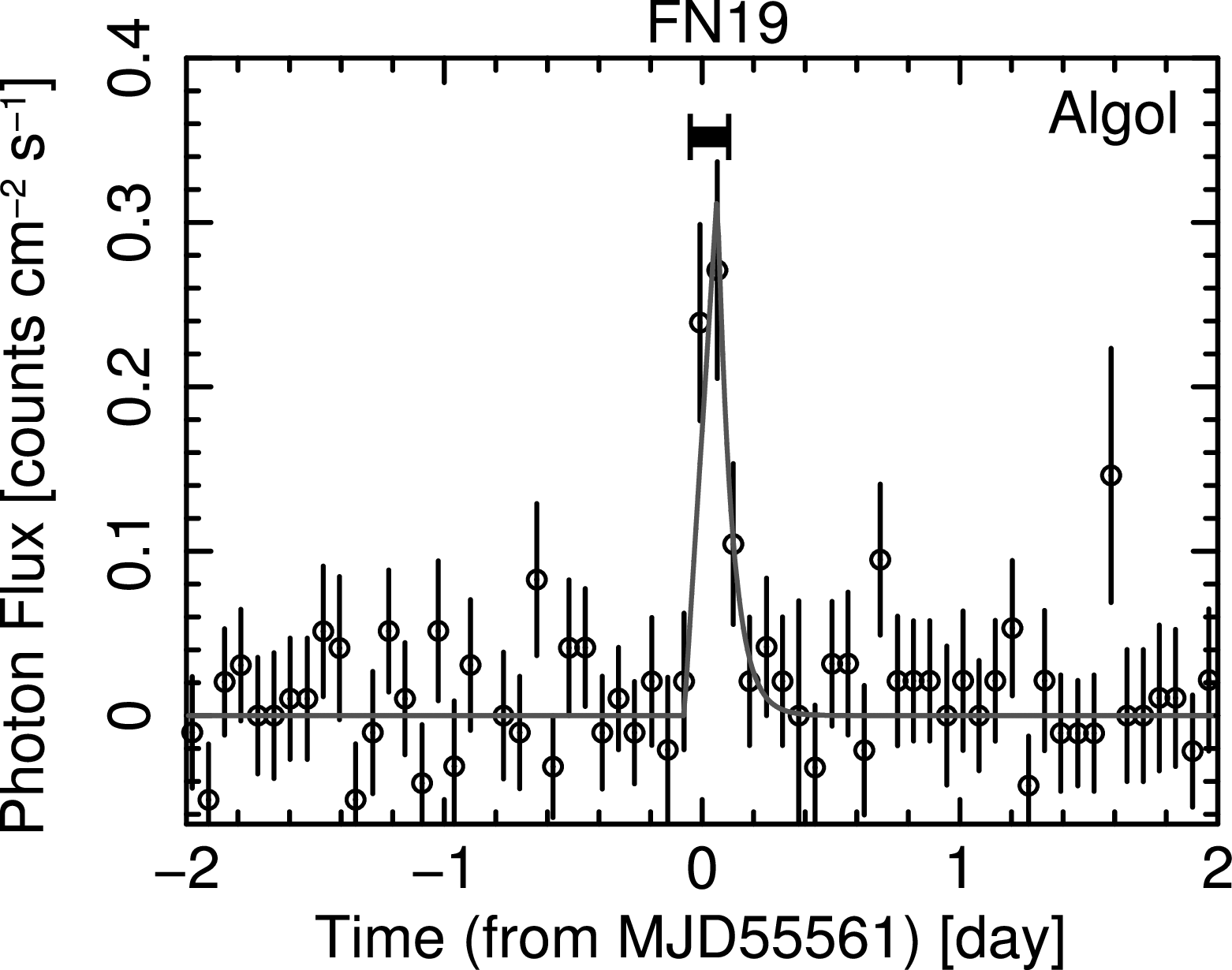}    
\end{center}
\end{minipage}
\end{figure*}
\clearpage
\begin{figure*}[htbp]
\begin{minipage}{0.5\hsize}
\begin{center}
\includegraphics[width=75mm]{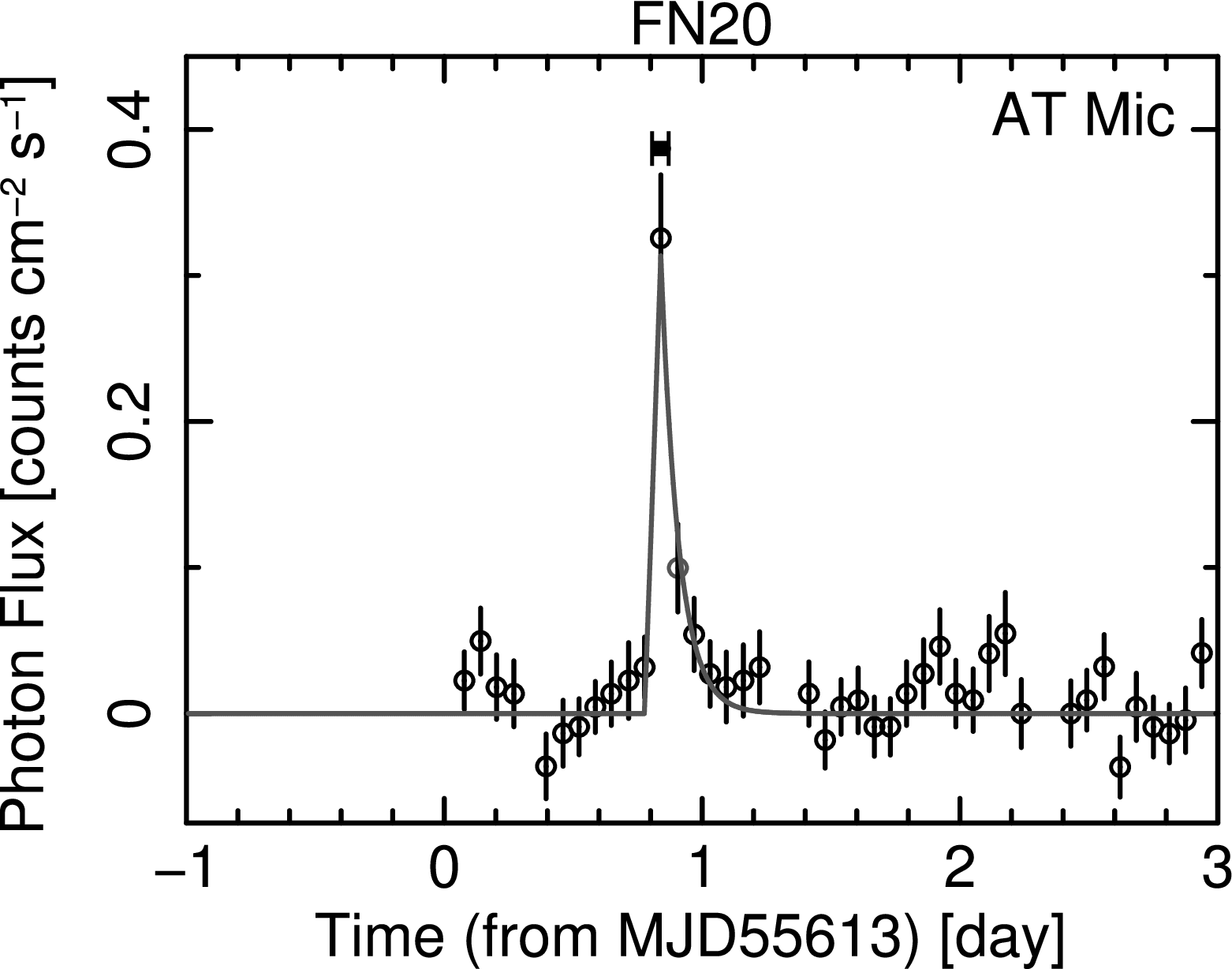}    
\end{center}
\end{minipage}
\begin{minipage}{0.5\hsize}
\begin{center}
\includegraphics[width=75mm]{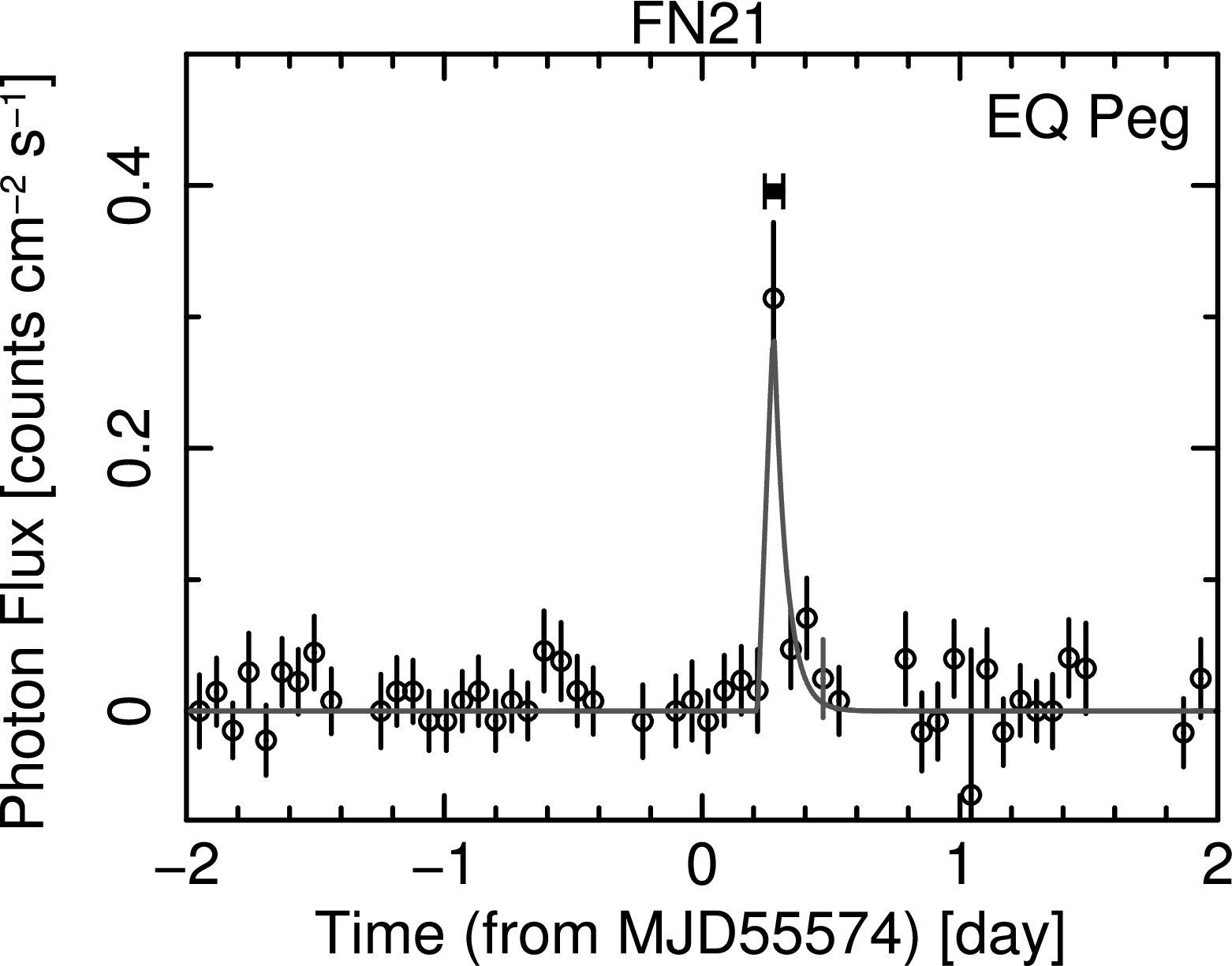}    
\end{center}
\end{minipage}
\begin{minipage}{0.5\hsize}
\begin{center}
\includegraphics[width=75mm]{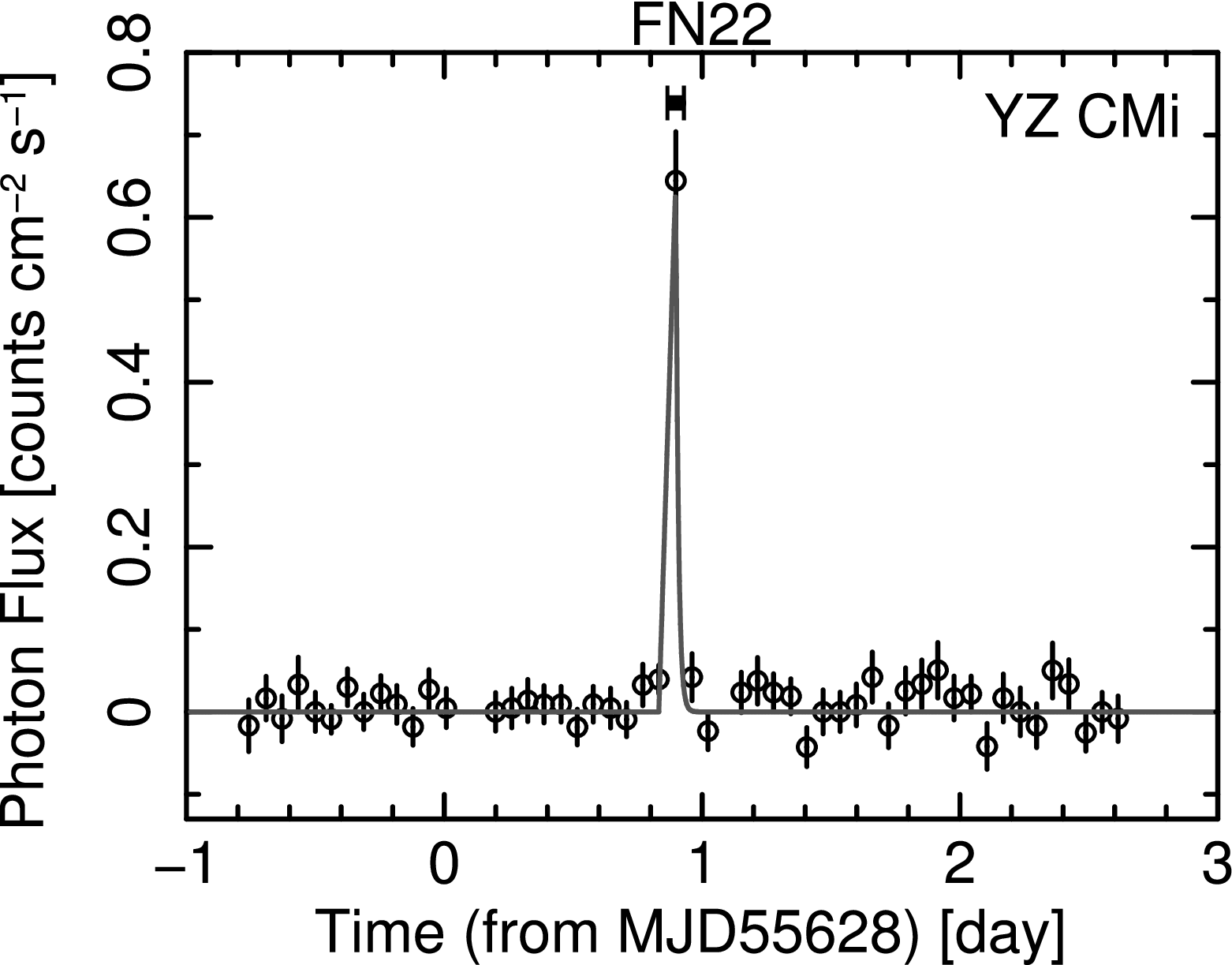}    
\end{center}
\end{minipage}
\begin{minipage}{0.5\hsize}
\begin{center}
\includegraphics[width=75mm]{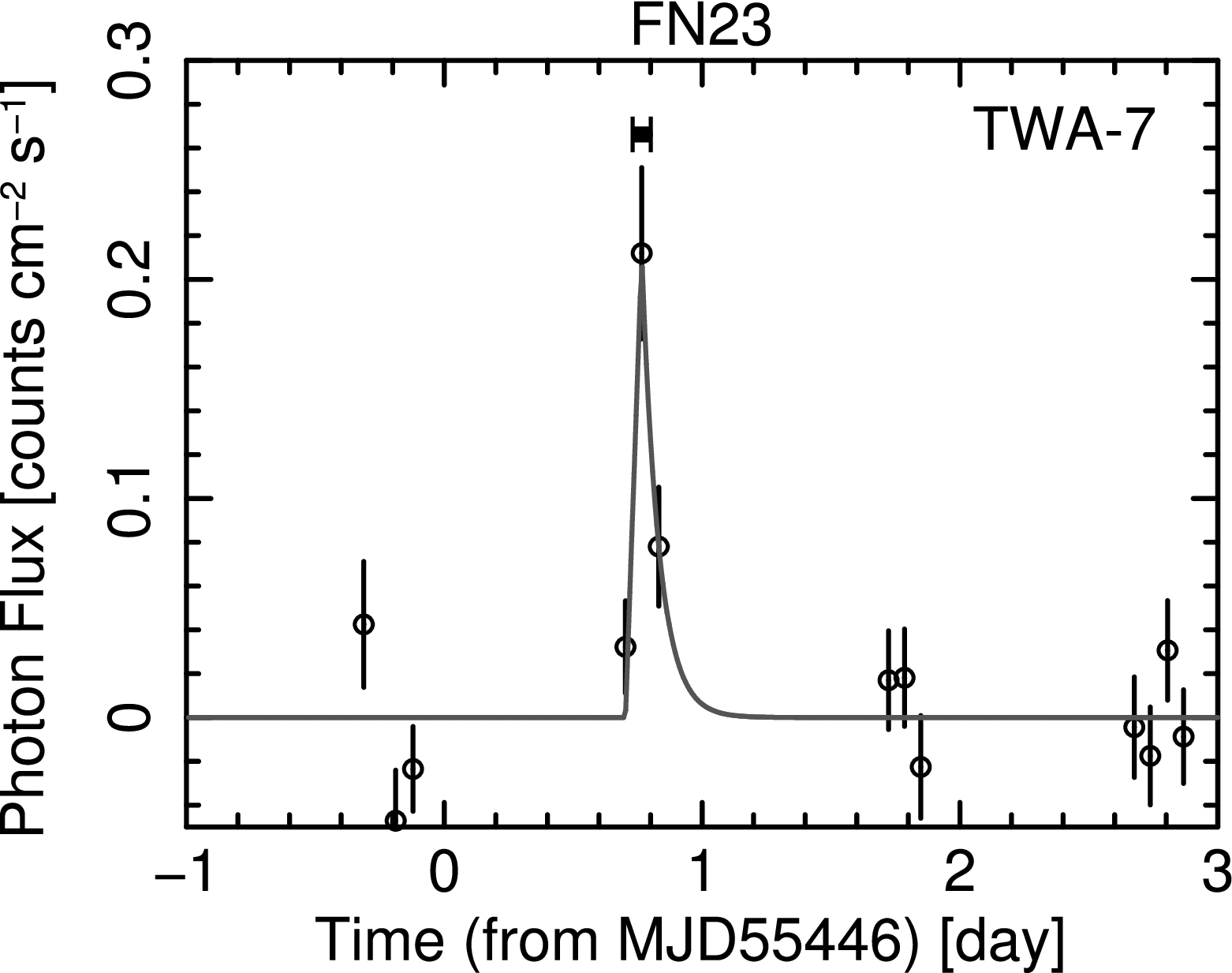}    
\end{center}
\end{minipage}
\caption{The light curves of twenty-three flares. The data are extracted in the
 2--10 keV band.  Each data-set shown by an open circle is made from one
 orbit data, while the datasets in the panels for FN1, 2, 9 and 14
 are binned with the data of multiple orbits.  The horizontal bar(s) 
  above the major line peak(s) in each panel is the time interval, from which the data are
 extracted to derive the detection significances and to make the
 spectral analysis.  The light curve of TWA-7 is from \citet{Uzawa+11}.}
 \label{lcurve}
\end{figure*}

\begin{figure*}[htbp]
\begin{center}
\includegraphics[width=100mm]{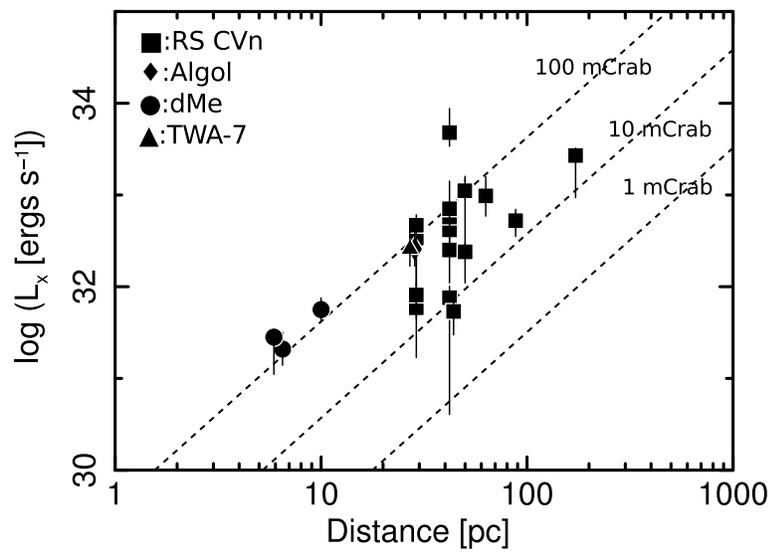}    
\end{center}
\caption{log-log plot of X-ray luminosity in the 2--20 keV band of
 flares vs. distance from active stars detected with MAXI/GSC. The
 filled squares, filled diamond, filled circles and filled triangle show
 RS-CVn type stars, Algol, dMe stars and TWA-7, respectively. The
 detection limit appeared to be roughly 10 mCrab in the 2--20 keV
 band.}\label{Dis_Lx}
\end{figure*}

\begin{figure*}[htbp]
\begin{minipage}{0.5\hsize}
\begin{center}
\includegraphics[width=80mm]{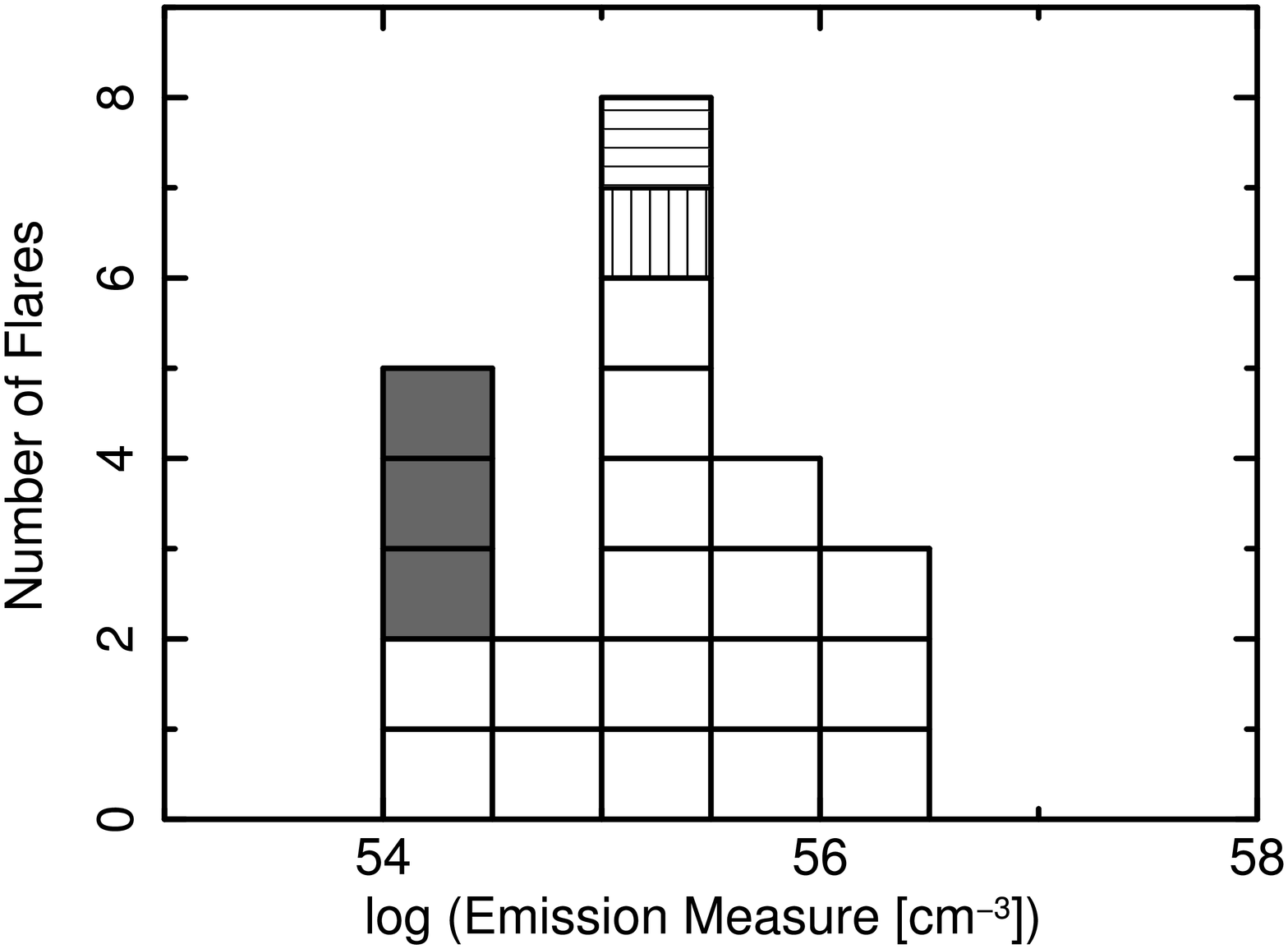}    
\end{center}
\end{minipage}
\begin{minipage}{0.5\hsize}
\begin{center}
\includegraphics[width=80mm]{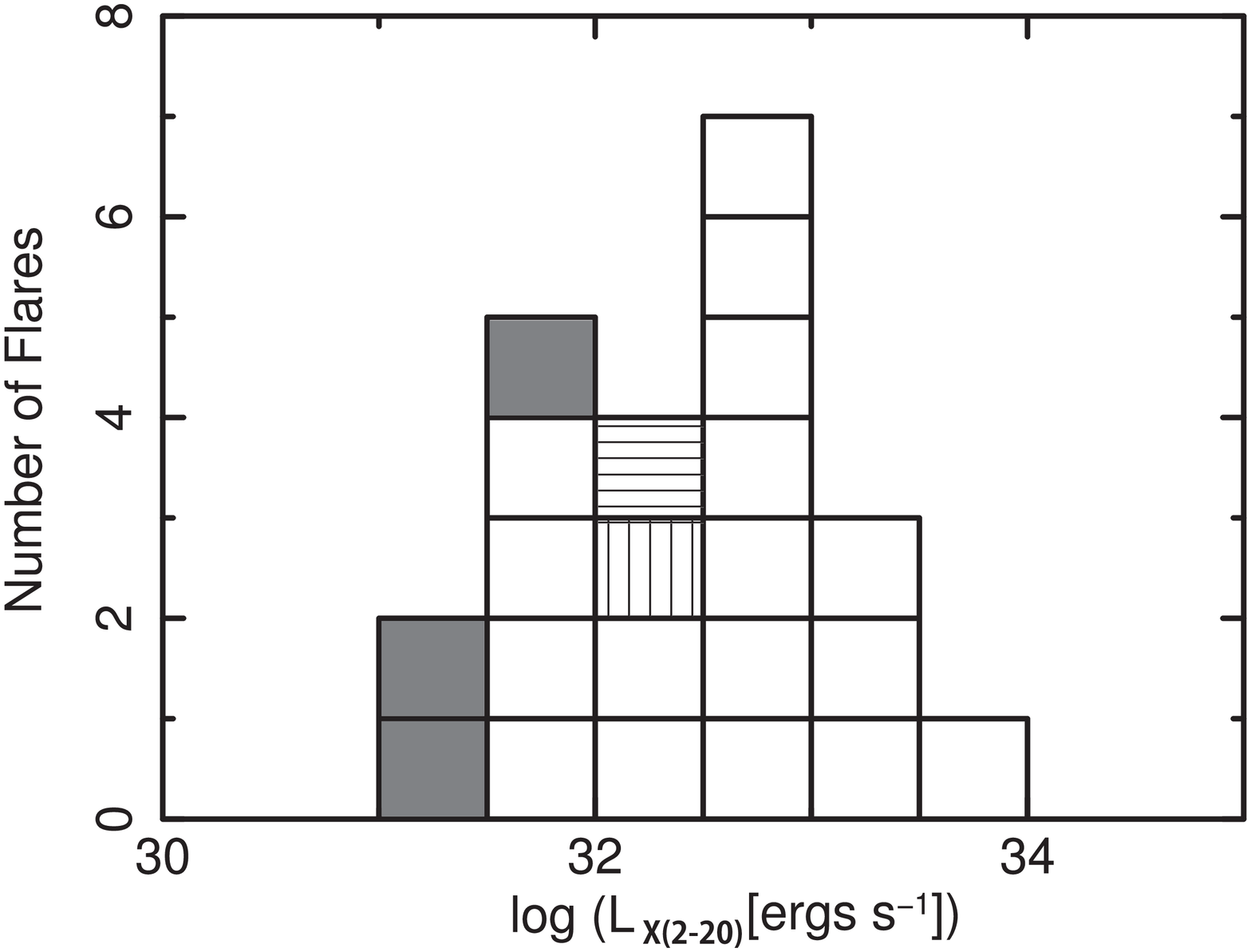}    
\end{center}
\end{minipage}
\begin{minipage}{0.5\hsize}
\begin{center}
\includegraphics[width=80mm]{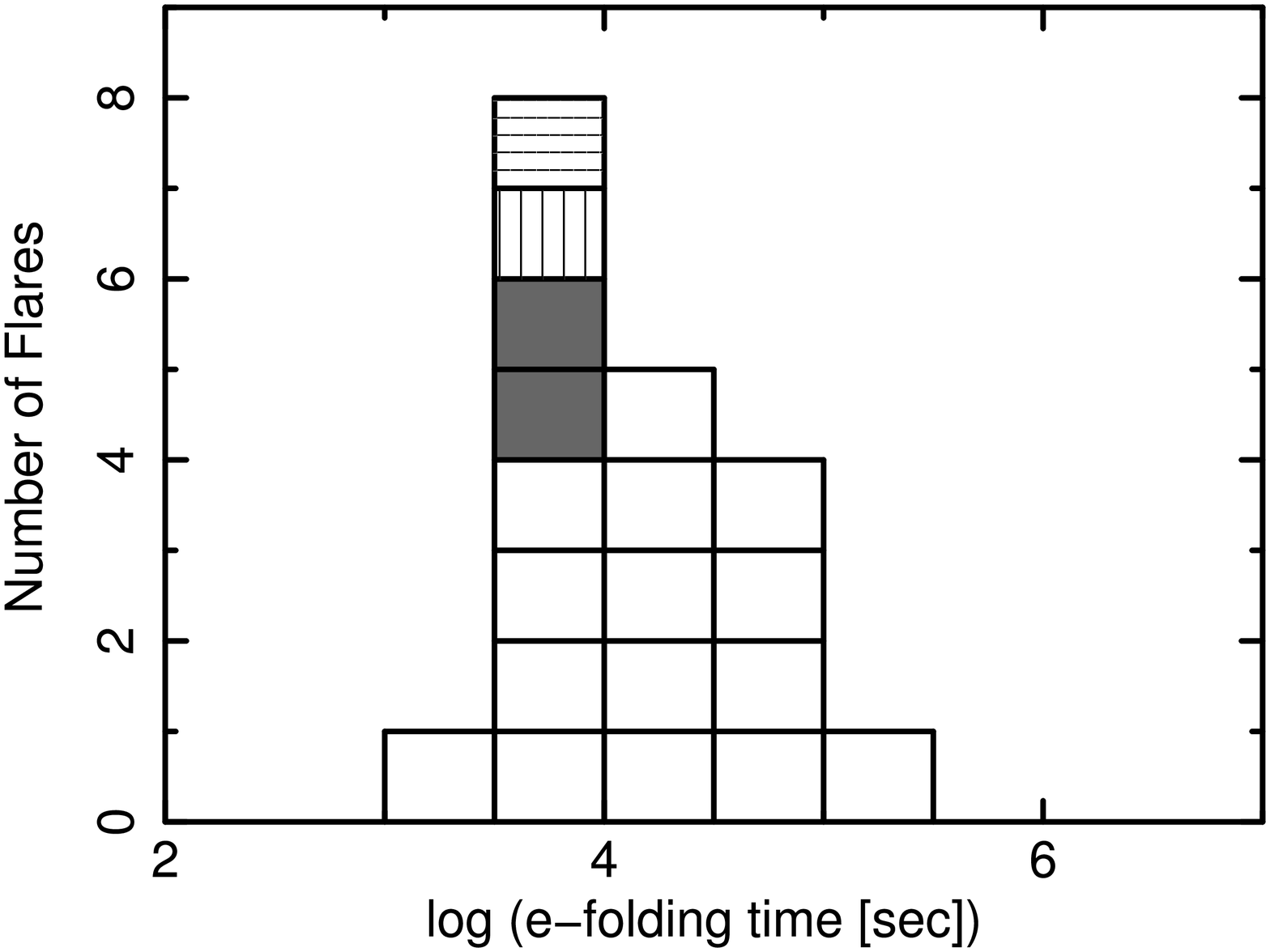}    
\end{center}
\end{minipage}
\begin{minipage}{0.5\hsize}
\begin{center}
\includegraphics[width=80mm]{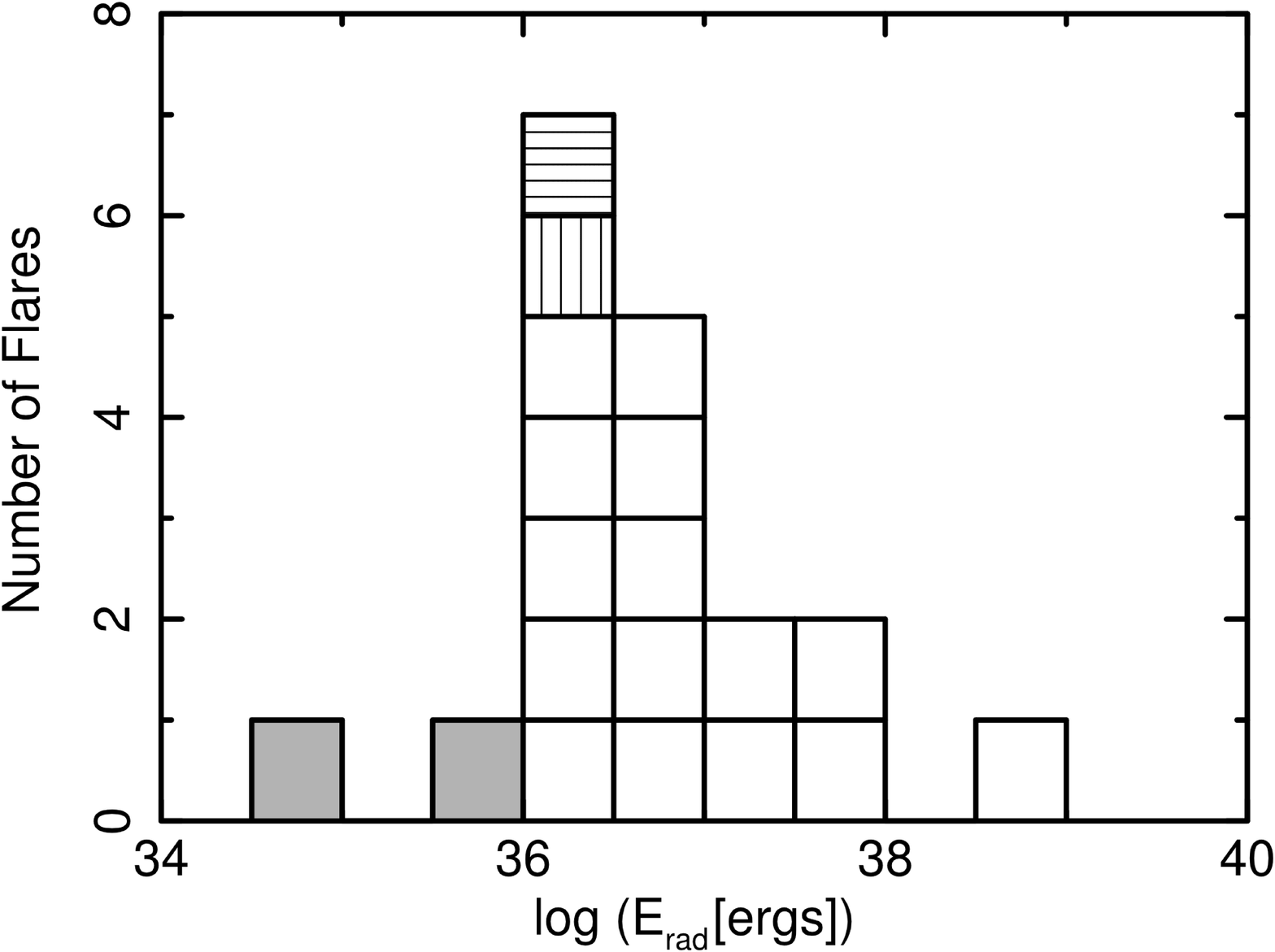}    
\end{center}
\end{minipage}
\caption{Distribution of the emission measure (top left), X-ray
 luminosity in the 2--20 keV band (top right), $e$-folding time
 (bottom left), and total energy (bottom right). The
 open squares, filled squares, vertical-striped square, and
 horizontal-striped square show RS-CVn type stars, dMe stars, Algol
 and TWA-7, respectively.}
\label{Histogram}
\end{figure*}

\begin{figure*}[htbp]
\begin{center}
\includegraphics[width=100mm]{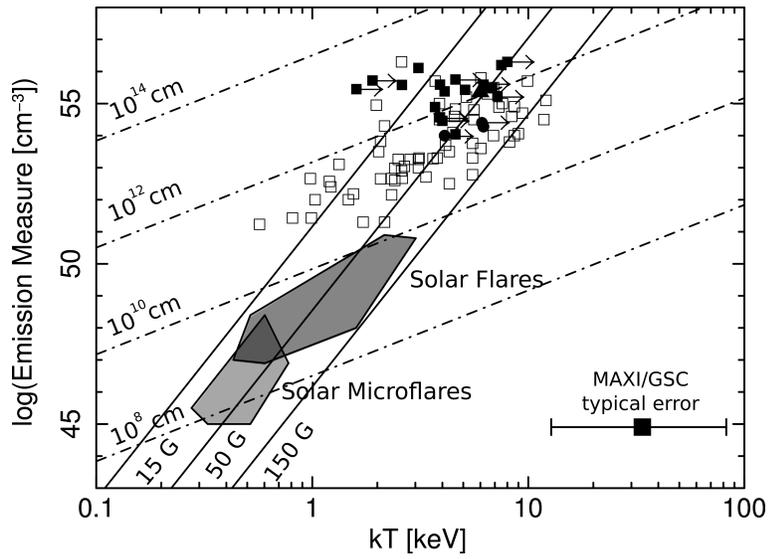}    
\end{center}
\caption{Log-log plot of emission measure
 ($EM$) vs. plasma temperature ($kT$) for the MAXI X-ray flares (filled symbols, as in figure~\ref{Dis_Lx}),  along with stellar flares from RS-CVn type, Algol, dMe stars and YSOs (open squares; the
 references  given in table \ref{ref_fig}), solar flares (\cite{Feldman+95}), and solar microflares (\cite{Shimizu95}). The
 arrows indicate the lower limits for individual MAXI/GSC sources.  The typical error
 for MAXI/GSC sources in 90\% confidence level is  indicated at the bottom-right corner.\newline The three solid lines are the theoretical $EM$-$kT$ relation,
 based on the equation [$EM \propto B^{-5}T^{17/2}$],  for $B$ = 15, 50, and 150 Gauss and the four dashed-dotted lines are that based on the
 equation [$EM \propto l^{5/3}T^{8/3}$] 
 for the loop-sizes of $10^8, 10^{10}$, $10^{12}$, and $10^{14}$ cm
 \citep{Shibata+99}.}  \label{kT_EM}
\end{figure*}


\begin{figure*}[htbp]
\begin{minipage}{\hsize}
\begin{center}
\includegraphics[width=120mm]{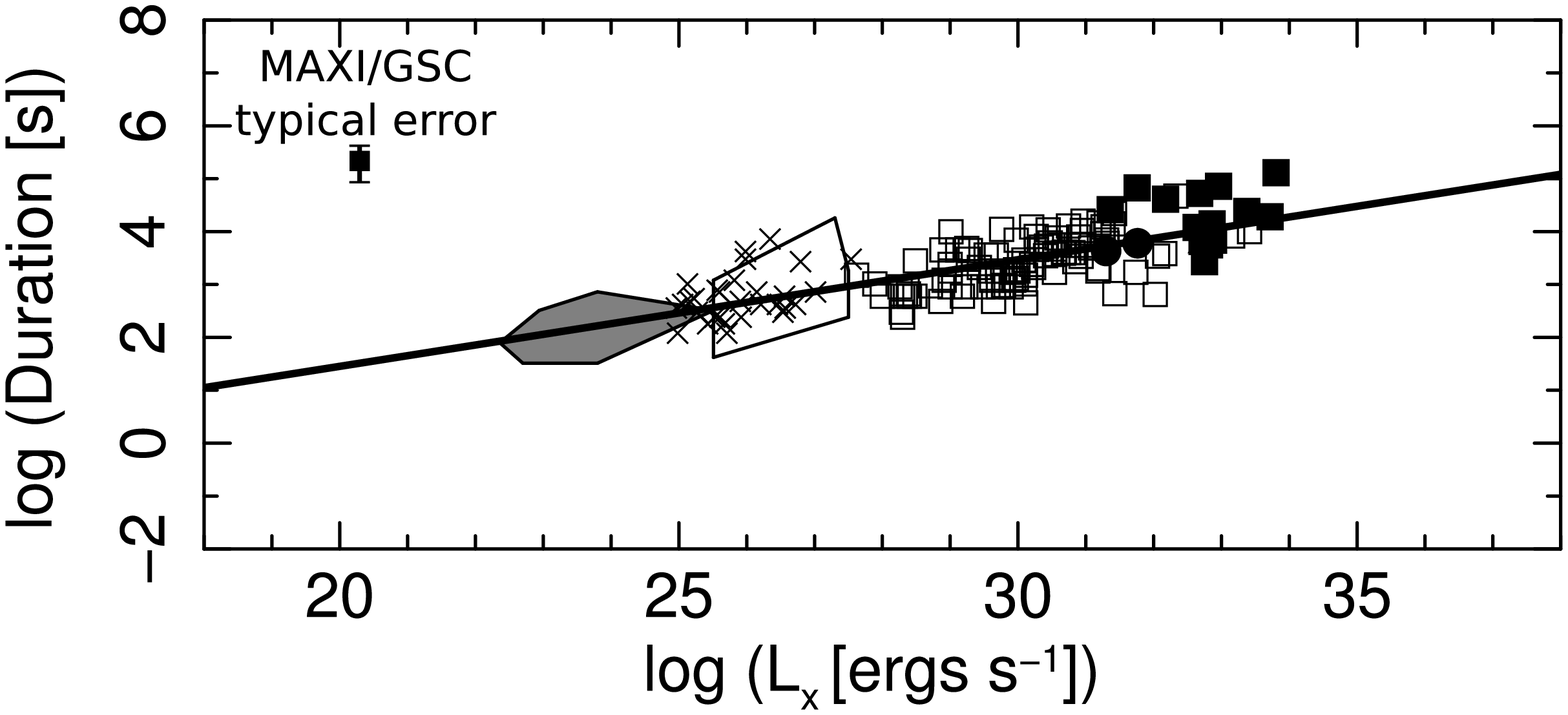}    
\end{center}
\end{minipage}
\begin{minipage}{\hsize}
\begin{center}
\includegraphics[width=120mm]{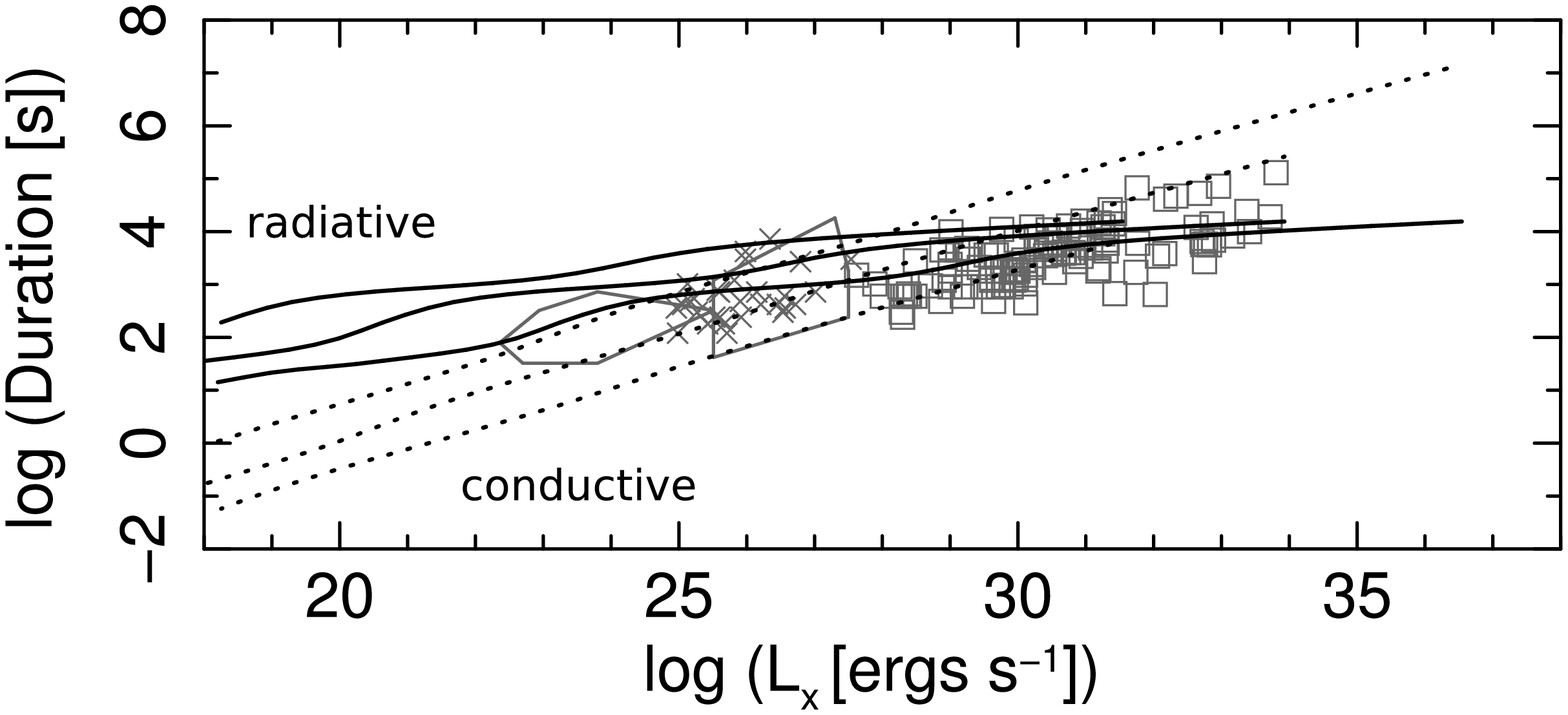}    
\end{center}
\end{minipage}
\begin{minipage}{\hsize}
\begin{center}
\includegraphics[width=120mm]{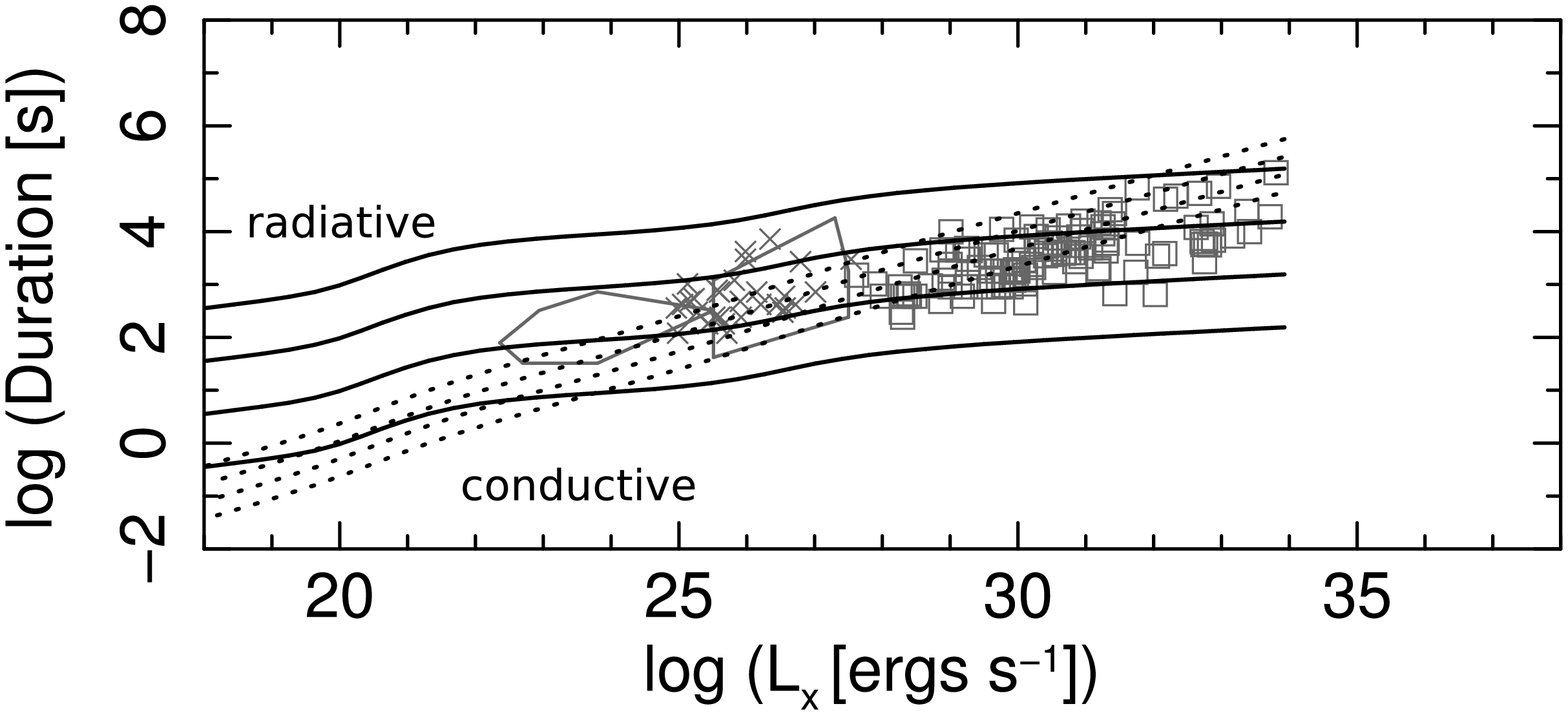}    
\end{center}
\end{minipage}
\caption{Top panel: Log-log plot of the duration of flares vs. X-ray
 luminosity in the 0.1--100 keV band. Our best-fit model is inserted
 with a broad solid line.  The filled and open symbols are the same as
 those in figure \ref{Dis_Lx} and \ref{kT_EM}, respectively. We
 superpose three sets of data of solar flares: X marks, large open
 pentagon and large gray region, taken from \citet{Pallavicini+77},
 \citet{Veronig+02} and \citet{Shimizu95}, respectively.  The typical
 error for MAXI/GSC sources is also inserted. Middle panel: The
 theoretical relations of the radiative cooling model (solid line) and conductive cooling model
 (dotted line) are  overlaid
 for the non-dimensional parameter $\alpha$ of 0.3, 1 and 3 from the top to bottom lines, respectively, for both the models, where   $n_{\rm e}$ is fixed to $10^{11}~{\rm
 cm}^{-3}$.  Bottom panel: The same as the middle panel, but  with  $\alpha$  fixed to 1 and $n_{\rm e}$ varied instead for  $n_{\rm e}~=$ $10^{10}$, $10^{11}$, $10^{12}$ and
 $10^{13}~{\rm cm}^{-3}$,  corresponding to the 4 lines from top to bottom, respectively, for each model.}  \label{Decay_Lx}
\end{figure*}

\begin{figure*}[htbp]
\begin{center}
\includegraphics[width=85mm]{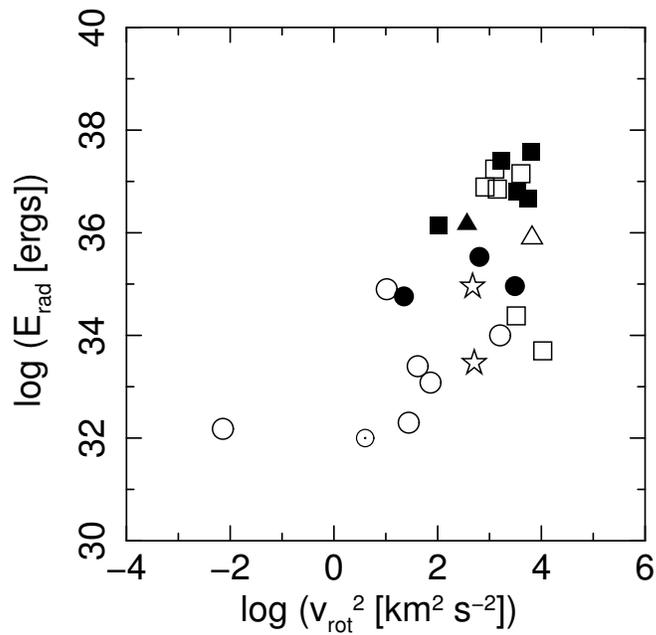}    
\end{center}
\caption{Log-log plot of total
energy released radiatively in a flare vs. the square of rotation velocity. The
filled symbols are the same as those in figure \ref{Dis_Lx}. The open
squares, open circles, open triangles, open stars and open circles with
a dot show the values obtained  in previous studies for RS-CVn type
stars, dMe stars, YSOs, dKe stars and the Sun respectively.
If flares have been detected from a source with MAXI or the other missions
more than once, only the largest $E_{\rm rad}$ is plotted. For the rotation
velocity, if the flare source is a multiple-star system, we used the
value of the component which has  the largest stellar radius.}  
\label{V_Etot}
\end{figure*}

\begin{figure*}[htbp]
\begin{center}
\includegraphics[width=80mm]{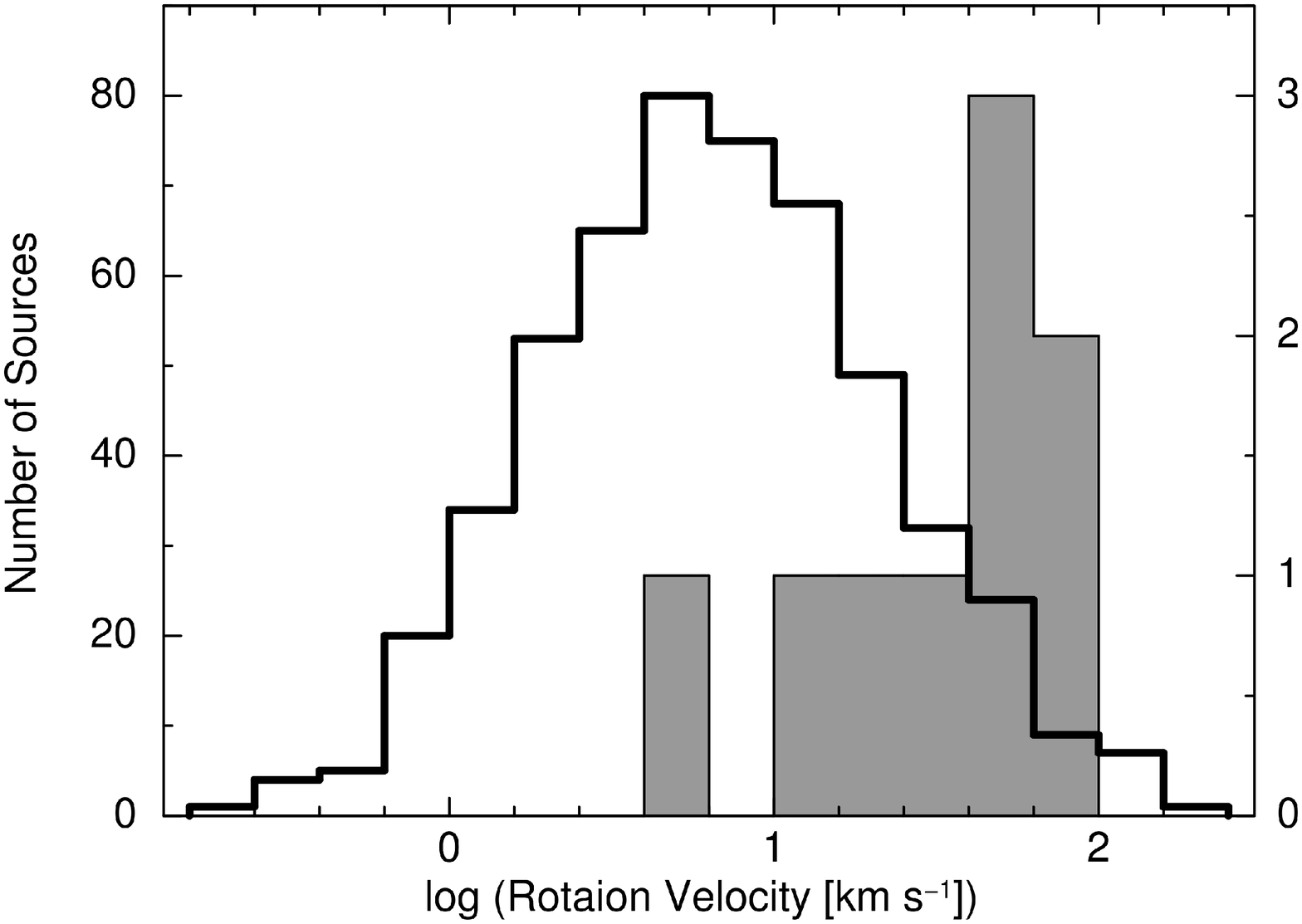}    
\end{center}
\caption{Histogram of rotation velocities. The gray bars indicate the
 MAXI/GSC-detected sources within 100 pc with the scale displayed on the right-hand Y-axis. The white bars indicate active
 binaries  and X-ray-detected stellar sources both within 100 pc
 (from \cite{Eker+08} and \cite{Wright+11}), with the scale displayed on the left-hand Y-axis, from which  the
 MAXI/GSC-detected sources are excluded.}  
\label{Histogram_V}
\end{figure*}

\begin{figure*}[htbp]
\begin{center}
\includegraphics[width=85mm]{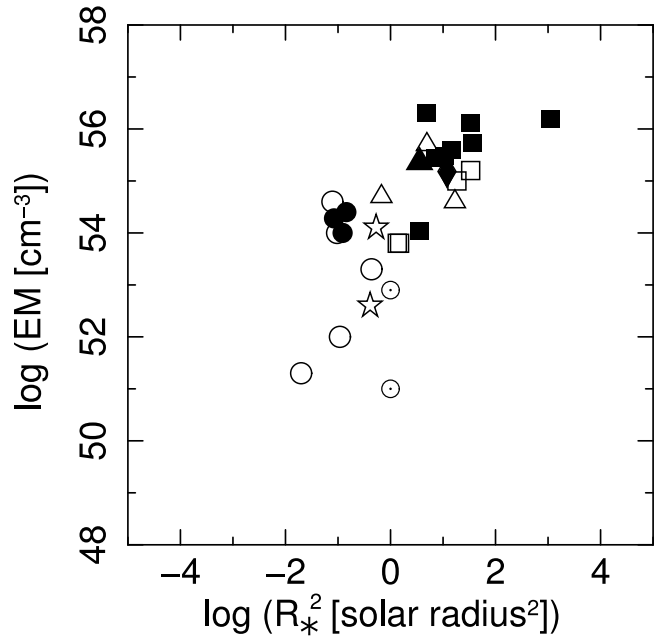}    
\end{center}
\caption{Log-log plot of emission measure vs. the square of radius.  
The symbols are the same as in  figure~\ref{V_Etot}.
The upper open circle with a dot indicates $\pi^{1}$ UMa, while the
lower does the Sun.}  \label{R_EM}
\end{figure*}

\begin{figure*}[htbp]
\begin{center}
\includegraphics[width=80mm]{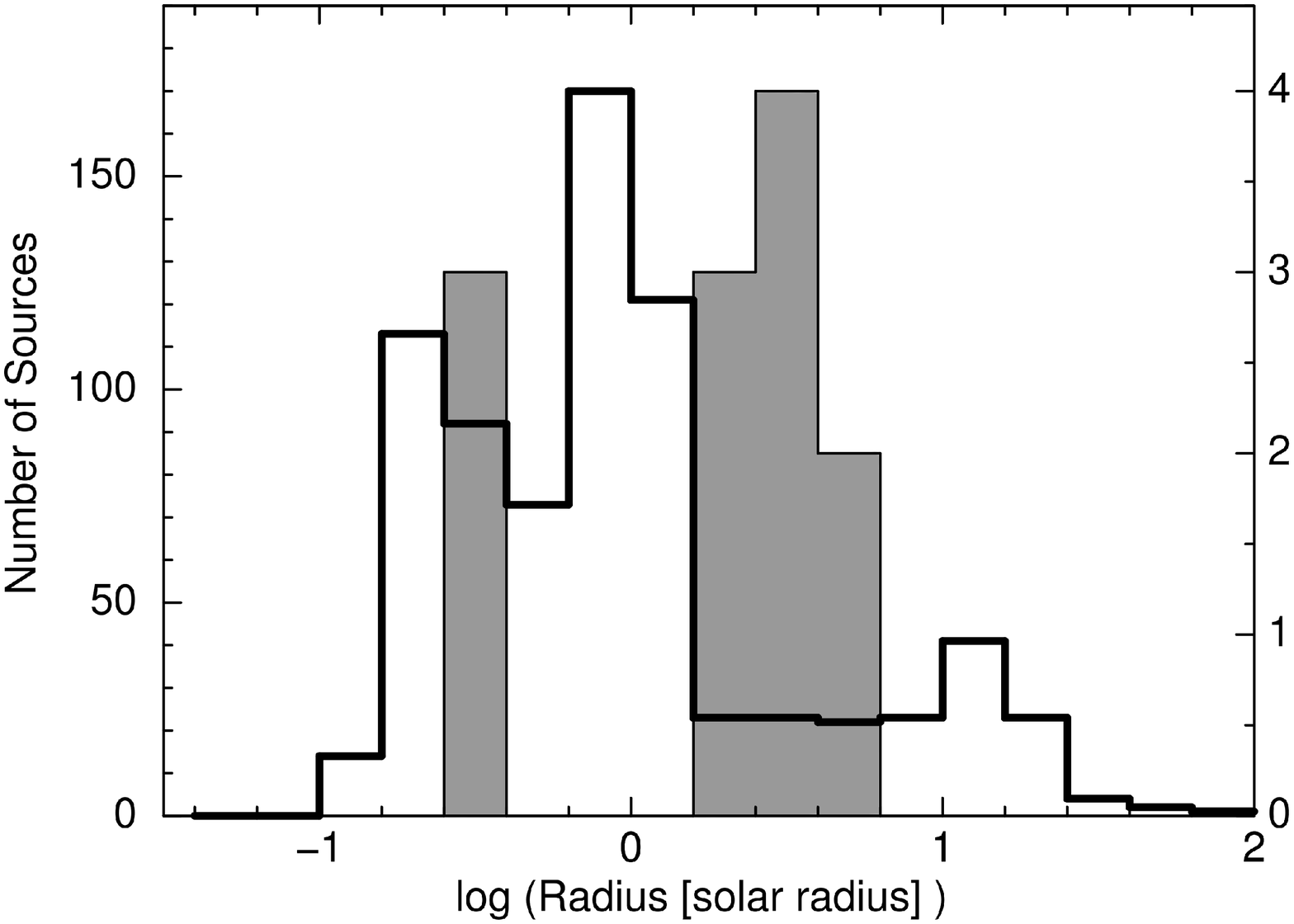}    
\end{center}
\caption{Histogram of stellar radii. The gray bars indicate the MAXI/GSC
sources within 100 pc with the scale displayed on the right-hand Y-axis. The white bars indicate active binaries  and X-ray-detected stellar sources both within 100 pc (from
\cite{Eker+08} and \cite{Wright+11}), with the scale displayed on the left-hand Y-axis, from which  the
MAXI/GSC-detected sources are excluded.}  \label{Histogram_R}
\end{figure*}



\begin{table*}[h]
\caption{References for figure \ref{kT_EM}, \ref{Decay_Lx}, \ref{V_Etot}, and \ref{R_EM}.}\label{ref_fig}
\begin{center}
{\scriptsize
\begin{tabular}{l|c||l|c}
\hline \hline
Reference & Figure & Reference & Figure \\
\hline
\cite{Alekseev+01} & \ref{V_Etot}&      \cite{Montmerle+83} & \ref{kT_EM} \\
\cite{Amado+00} & \ref{R_EM}&   \cite{Morales+09} & \ref{R_EM} \\           
\cite{Anders+99} & \ref{V_Etot}&        \cite{Morin+08} & \ref{V_Etot},\ref{R_EM} \\
\cite{Benedict+98} & \ref{V_Etot}&      \cite{Murdoch+95} & \ref{R_EM} \\           
\cite{Bopp+89} & \ref{R_EM}&    \cite{OBrien+01} & \ref{V_Etot},\ref{R_EM} \\       
\cite{Briggs+03} & \ref{kT_EM}& \cite{ONeal+01} & \ref{V_Etot},\ref{R_EM} \\        
\cite{Covino+01} & \ref{kT_EM},\ref{V_Etot},\ref{R_EM}& \cite{Osten+98} & \ref{V_Etot} \\
\cite{Demory+09} & \ref{V_Etot},\ref{R_EM}&     \cite{Osten+07} & \ref{Decay_Lx} \\      
\cite{Dempsey+93} & \ref{V_Etot}&       \cite{Osten+10} & \ref{kT_EM},\ref{Decay_Lx},\ref{V_Etot},\ref{R_EM} \\
\cite{Donati99} & \ref{R_EM}&   \cite{Ottmann+94} & \ref{kT_EM} \\                                             
\cite{Donati+97} & \ref{V_Etot}&        \cite{Ozawa+99} & \ref{kT_EM} \\                                       
\cite{Doyle+88} & \ref{kT_EM}&  \cite{Pallavicini+90a} & \ref{kT_EM},\ref{Decay_Lx},\ref{V_Etot},\ref{R_EM} \\ 
\cite{Duemmler+01} & \ref{V_Etot},\ref{R_EM}&   \cite{Pallavicini+90b} & \ref{kT_EM} \\                        
\cite{Eaton+07} & \ref{V_Etot},\ref{R_EM}&      \cite{Pan+95} & \ref{kT_EM},\ref{V_Etot},\ref{R_EM} \\
\cite{Endl+97} & \ref{kT_EM},\ref{V_Etot},\ref{R_EM}&   \cite{Pan+97} & \ref{kT_EM},\ref{Decay_Lx},\ref{R_EM} \\
\cite{ESA97} & \ref{Decay_Lx},\ref{V_Etot}&     \cite{Pandey+08} & \ref{kT_EM},\ref{Decay_Lx},\ref{V_Etot},\ref{R_EM} \\
\cite{Favata+99} & \ref{kT_EM}& \cite{Pettersen80} & \ref{V_Etot} \\
\cite{Favata+00a} & \ref{Decay_Lx}&     \cite{Pettersen89} & \ref{R_EM} \\
\cite{Favata+00b} & \ref{kT_EM},\ref{V_Etot},\ref{R_EM}&        \cite{Pizzolato+03} & \ref{V_Etot} \\
\cite{Favata+01} & \ref{kT_EM},\ref{V_Etot},\ref{R_EM}& \cite{Poletto+88} & \ref{kT_EM} \\
\cite{Fekel+99} & \ref{V_Etot},\ref{R_EM}&      \cite{Preibisch+93} & \ref{kT_EM} \\
\cite{Franciosini+01} & \ref{kT_EM},\ref{Decay_Lx}&     \cite{Preibisch+95} & \ref{kT_EM} \\
\cite{Frasca+97} & \ref{V_Etot}&        \cite{Pribulla+01} & \ref{V_Etot},\ref{R_EM} \\
\cite{Gudel+99} & \ref{kT_EM}&  \cite{Pye+83} & \ref{Decay_Lx},\ref{V_Etot} \\
\cite{Gudel+04} & \ref{kT_EM},\ref{V_Etot},\ref{R_EM}&  \cite{Qian+02} & \ref{R_EM} \\
\cite{Gagne+95} & \ref{kT_EM}&  \cite{Ramseyer+95} & \ref{V_Etot} \\
\cite{Glebocki+95} & \ref{V_Etot}&      \cite{Randich+93} & \ref{V_Etot} \\
\cite{Gunn+98} & \ref{R_EM}&    \cite{Reiners+07} & \ref{V_Etot} \\
\cite{Hamaguchi+00} & \ref{kT_EM}&      \cite{Reiners+09} & \ref{V_Etot} \\
\cite{Hatzes95} & \ref{V_Etot}& \cite{Robinson+03} & \ref{V_Etot} \\
\cite{Huensch+95} & \ref{kT_EM}&        \cite{Sanz+03} & \ref{V_Etot},\ref{R_EM} \\
\cite{Hussain+05} & \ref{V_Etot}&       \cite{Singh+95} & \ref{V_Etot} \\
\cite{Imanishi+01} & \ref{kT_EM}&       \cite{Stawikowski+94} & \ref{V_Etot} \\
\cite{Imanishi+03} & \ref{Decay_Lx}&    \cite{Stern+83} & \ref{kT_EM} \\
\cite{Jeffries+90} & \ref{kT_EM},\ref{V_Etot},\ref{R_EM}&       \cite{Stern+92} & \ref{Decay_Lx} \\
\cite{Kahler+82} & \ref{kT_EM},\ref{V_Etot}&    \cite{Strassmeier+93} & \ref{V_Etot} \\
\cite{Kamata+97} & \ref{kT_EM}& \cite{Strassmeier+94} & \ref{V_Etot},\ref{R_EM} \\
\cite{Kjurkchieva+00} & \ref{V_Etot}&   \cite{Strassmeier+98} & \ref{V_Etot},\ref{R_EM} \\
\cite{Kovari+01} & \ref{V_Etot}&        \cite{Strassmeier+03} & \ref{V_Etot} \\
\cite{Kuerster+96} & \ref{kT_EM},\ref{V_Etot},\ref{R_EM}&       \cite{Torres+02} & \ref{V_Etot},\ref{R_EM} \\
\cite{Landini+86} & \ref{kT_EM},\ref{R_EM}&     \cite{Tsuboi+98} & \ref{kT_EM},\ref{Decay_Lx},\ref{R_EM} \\
\cite{Lim+87} & \ref{R_EM}&     \cite{Tsuboi+00} & \ref{kT_EM} \\
\cite{Linsky91} & \ref{Decay_Lx}&       \cite{Tsuru+89} & \ref{kT_EM} \\
\cite{Linsky+01} & \ref{R_EM}&  \cite{Oord+88} & \ref{kT_EM},\ref{V_Etot},\ref{R_EM} \\
\cite{Maggio+00} & \ref{kT_EM},\ref{Decay_Lx},\ref{V_Etot},\ref{R_EM}&  \cite{Oord+89} & \ref{V_Etot},\ref{R_EM}\\
\cite{Mewe+97} & \ref{kT_EM}&   \cite{Welty95} & \ref{V_Etot},\ref{R_EM} \\
\cite{Miranda+07} & \ref{R_EM}& \cite{White+94} & \ref{R_EM} \\
\cite{Mitra07} & \ref{V_Etot}&  \cite{Wright+11} & \ref{V_Etot},\ref{R_EM} \\
\cite{Mitra07} & \ref{V_Etot}&  \cite{Yang+08} & \ref{V_Etot},\ref{R_EM} \\
\cite{Miura+08} & \ref{Decay_Lx}&       \cite{Zboril+05} & \ref{V_Etot},\ref{R_EM} \\
\cite{Montes+95} & \ref{V_Etot}& &\\
\hline
\end{tabular} }
\end{center}
\end{table*}

\end{document}